\documentclass[hyperref,letterpaper,floatfix,reprint,aps,pre,superscriptaddress,longbibliography]{revtex4-1}
\pdfoutput=1

\usepackage{graphicx}
\usepackage{hyperref}

\usepackage{amssymb,amsmath,amsfonts}
\usepackage{times}
\usepackage{array}
\usepackage{tabularx}
\usepackage{placeins}
\newcolumntype{M}{>{$\vcenter\bgroup\hbox\bgroup}c<{\egroup\egroup$}}

\renewcommand{\topfraction}{1.00}
\renewcommand{\dbltopfraction}{1.00}
\renewcommand{\textfraction}{0.00}
\renewcommand{\floatpagefraction}{0.5}
\renewcommand{\dblfloatpagefraction}{0.5}

\usepackage{ifthen}
\usepackage{verbatim}

\usepackage{youngtab}
\usepackage{tikz}
\usetikzlibrary{arrows,shapes,positioning,calc,3d,shapes.geometric}
\usepackage{makecell}

\usepackage{color}

\newcommand{\Zcal}{\mathcal{Z}}

\newcommand{\via}{\textit{via}}
\newcommand{\eg}{\textit{e.g.}}
\newcommand{\ie}{\textit{i.e.}}
\newcommand{\cf}{\textit{c.f.}}

\newcommand{\vs}{\textit{vs.}}

\newcommand{\SI}{appendix}

\DeclareMathOperator{\sgn}{sgn}

\usepackage[english]{babel} 
\usepackage[protrusion=true,expansion=true]{microtype} 
\usepackage{booktabs} 

\definecolor{mbd}{HTML}{00274C}
\definecolor{mbl}{HTML}{587ABC}
\definecolor{mmaize}{HTML}{FFCB05}

\definecolor{gcb1}{cmyk}{0.35,0.07,0,0}
\definecolor{gcb2}{cmyk}{0.90,0.30,0,0}
\definecolor{gcb3}{cmyk}{0.30,0,0.45,0}
\definecolor{gcb4}{cmyk}{0.80,0,1.00,0}
\definecolor{gcb5}{cmyk}{0,0.40,0.25,0}
\definecolor{gcb6}{cmyk}{0.10,0.90,0.80,0}
\definecolor{gcb7}{cmyk}{0,0.25,0.50,0}
\definecolor{gcb8}{cmyk}{0,0.50,1.00,0}
\definecolor{gcb9}{cmyk}{0.20,0.25,0,0}
\definecolor{gcb10}{cmyk}{0.60,0.70,0,0}

\begin{document}
\title{Digital Alchemy for Materials Design: Colloids and Beyond}
\author{Greg \surname{van Anders}}
\affiliation
{Department of Chemical Engineering, University of Michigan, Ann Arbor,
MI 48109-2136, USA}
\author{Daphne \surname{Klotsa}}
\affiliation
{Department of Chemical Engineering, University of Michigan, Ann Arbor,
MI 48109-2136, USA}
\affiliation
{School of Engineering and Applied Sciences,
Harvard University, Cambridge, Massachusetts 02138, USA}
\affiliation
{Department of Chemistry, University of Cambridge, Lensfield Road,
Cambridge CB2 1EW, UK}
\author{Andrew S.\ Karas}
\affiliation
{Department of Chemical Engineering, University of Michigan, Ann Arbor,
MI 48109-2136, USA}
\author{Paul M.\ Dodd}
\affiliation
{Department of Chemical Engineering, University of Michigan, Ann Arbor,
MI 48109-2136, USA}
\author{Sharon C.\ Glotzer}
\affiliation
{Department of Chemical Engineering, University of Michigan, Ann Arbor,
MI 48109-2136, USA}
\affiliation
{Department of Materials Science and Engineering, University of Michigan, Ann
Arbor, MI 48109-2136, USA}
\affiliation{Biointerfaces Institute,
University of Michigan, Ann Arbor, MI 48109-2800, USA}
\begin{abstract}
  Starting with the early alchemists, a holy grail of science has been to make
  desired materials by modifying the attributes of basic building blocks.
  Building blocks that show promise for assembling new complex materials can be
  synthesized at the nanoscale with attributes that would astonish the ancient
  alchemists in their versatility. However, this versatility means that making
  direct connection between building block attributes and bulk behavior is both
  necessary for rationally engineering materials, and difficult because building
  block attributes can be altered in many ways. Here we show how to exploit the
  malleability of the valence of colloidal nanoparticle ``elements'' to directly
  and quantitatively link building block attributes to bulk behavior through a
  statistical thermodynamic framework we term ``digital alchemy''. We use this
  framework to optimize building blocks for a given target structure, and to
  determine which building block attributes are most important to control for
  self assembly, through a set of novel thermodynamic response functions, moduli
  and susceptibilities. We thereby establish direct links between the attributes
  of colloidal building blocks and the bulk structures they form. Moreover, our
  results give concrete solutions to the more general conceptual challenge of
  optimizing emergent behaviors in nature, and can be applied to other types of
  matter. As examples, we apply digital alchemy to systems of truncated
  tetrahedra, rhombic dodecahedra, and isotropically interacting spheres that
  self assemble diamond, FCC, and icosahedral quasicrystal structures,
\end {abstract}
\maketitle

Mendeleev's tabular organization of the
elements\cite{mendeleev1871,mendeleev1889} by atomic valence \cite{lewisvalence}
has served for more than 140 years as a heuristic that relates properties of the
atomic elements to how they arrange in bulk structures. However, attempts to
understand how properties of bulk structures relate to atomic properties predate
Mendeleev and, in fact, modern science \cite{makersofchem}, and are complicated
by the fact that the chemical manipulation of atoms is prohibited by the
quantization of both electrical charge and angular momentum. Fortunately for
Mendeleev, this quantization constrains Nature to only about 80 stable elements,
and limits elemental properties and bulk behaviors so that the elements can be
tabulated by valence. In fact, starting with technetium \cite{element43} in the
1930s, new atomic elements have only been produced artificially (as
suggested by the etymology of the name ``technetium'' \cite{technetium}) by
$\alpha$-particle bombardment, fusion, or other nuclear techniques that finally
realized the ancient alchemists' goal of transmuting the elements.

In contrast, an inexhaustible array of new ``elements'' can be synthesized
as patchy particles.\cite{patchy,glotzsolomon}
However,
the exploding diversity of patchy particles
\cite{glotzsolomon,pawarkretzchmar,shapecolloids} or, more generally,
colloidal ``elements'' means that there are now so
many types to synthesize and study that synthesizing them all and determining
their bulk behavior is no longer possible in practice. This fundamental
impracticality means that, for colloid science to progress, scientists must
first ask and answer the basic but daunting question ``What elements should I
make?'' Materials science that starts with this question must be carried out in
a fundamentally different way than traditional approaches, guided by the
question: What is the optimal building block to make for a given structure, and
why is it optimal?

Constructing a periodic table of colloidal elements is easier said than done,
however, because, unlike for atoms, colloid valence
\cite{zhangetal2,colloidmolecule,pinedna,epp,gangdirbind} is not discrete.
Moreover, entropic colloid valence \cite{epp,entint} is a collective effect
\cite{moreisdifferent} that emerges only when colloids are crowded \cite{epp,entint}.
A first step in constructing a periodic table of colloidal elements was taken by
heuristically classifying building blocks according to their valence along
anisotropy dimensions \cite{glotzsolomon,epp} that systematically and
orthogonally vary colloidal element attributes. This sort of colloidal alchemy
is now possible.

Here we present a statistical thermodynamic framework that forms the basis for a
new computational approach to building block design, which we term digital
alchemy.
Using this framework:
(i)~We show how to
treat anisotropy dimensions \cite{glotzsolomon,epp} or other particle
interaction parameters as thermodynamic variables, and interpret their conjugate
quantities. Treating particle interaction parameters thermodynamically means
that the attributes of the colloidal ``elements'' we study can change
so we refer to our methods as ``alchemy''
The term alchemy has been used previously in the modern era in the
context of materials design, and these uses are either different in spirit from
the present work \cite{sandhage}, or are focused on computing global free energy
differences in systems \cite{teter,shirtsalch} in which intermediate state
points are unphysical. A related investigation was also carried out in
Ref.\ \cite{batista}, which considered the effects of non-rigid colloid shape on
crystallization mechanically, whereas here we study rigid colloids that
fluctuate thermally. Though there are many systematic investigations of how
particle shape or interactions affect structure
\cite{dzugutov1,dijkstramix,michaelljg,carolynljg,pacman,dijkstrabowl,
dijkstrabowl2,dijkstranonconvex,rossi,gang11,shapecolloids,kraftetal,
trunctet,dijkstrasuperballs,zoopaper,stefanosynth,dijkstratcube,truskettsm,
epp,lockkey3d,stickydimple,andresshapemod,opp,sphinx}, we are aware of no work
that attempts to directly probe the thermodynamic response of a system to a
change in the attributes of its constituent building blocks.
in analogy with pre-scientific attempts to modify chemical
elements.\cite{makersofchem}
(ii)~We show how constitutive relations between anisotropy parameters and the
thermodynamically conjugate variables we term ``alchemical potentials'' encode
a broad class of detailed quantitative relations between building block
attributes and bulk behavior. Further, we define new moduli and susceptibilites
that describe stress-strain relationships between bulk structure and particle
attributes.
(iii)~We show that these building block \vs\ bulk relationships persist
in systems with entropy-driven, emergent collective behavior.
(iv)~We show how building block \vs\ bulk relationships can be used
both to determine optimal particle shapes or interactions for given structures,
and to compute the relative importance of different particle attributes for bulk
behavior.
(v)~We report a detailed, general microscopic design rule for a macroscopic,
entropy-driven, emergent behavior.
(vi)~We demonstrate this design rule in simulations that allow particle shape to
fluctuate dynamically by showing that when particles are constrained to sit on a
target lattice, they spontaneously adopt their preferred shape; that is, the
shape that minimizes the free energy of the target structure at a given state
point.

Through all of these findings we demonstrate what the outlines of a periodic
table of colloidal elements might look like. In particular, because colloidal
valence is not constrained, a colloidal periodic table cannot be as succinct as
the atomic periodic table. However, because colloidal valence can be
manipulated, a colloidal periodic table can encode detailed, quantitative
relationships between building block attributes and bulk behavior, and can tell
us what building blocks are optimal for a given structure, and why they are
optimal.

\section{Theoretical Results}
We consider a family of colloidal elements that can be described by a set of
isotropic interaction potentials, or by anisotropy dimensions for enthalpic
\cite{glotzsolomon} or entropic \cite{epp} patches, with parameters
$\{\alpha_i\}$.
The particles are described by a classical Hamiltonian $H$ that depends
on the $\alpha_i$ \via\ a pair interaction between particles, and the
rotational kinetic term in the Hamiltonian
\begin{equation}\label{hamiltonian}
  H(\{\alpha_i\}) = \frac{p^2}{2m}
  + \frac12 L^T I_{\{\alpha_i\}}^{-1} L + U_{\{\alpha_i\}}(q,Q) \; ,
\end{equation}
where $p$ are momenta, $L$ are angular momenta, $I$ is the moment of inertia
tensor, and $U$ is the interaction potential that depends on particle positions
$q$ and orientations $Q$, and where we have suppressed particle indices.
We consider systems in which the generalized particle coordinates and their
conjugate momenta do not have explicit dependence on the $\alpha_i$. In
this case, the $\alpha_i$ have vanishing Poisson brackets with the
Hamiltonian, and are invariants of the system: $\{ \alpha_i, H\} =0$. This is
the case if, \eg, a particle's shape is independent of its generalized momentum
and position. This would not hold, \eg, for systems with chemical gradients that
cause a particle to swell in some locations more than others. Furthermore, we
consider systems in which the $\alpha_i$ themselves are mutually commuting, \ie,
the order in which operations are applied to modify the building blocks is not
important. We regard the $\alpha_i$ as a set of mutually conserved charges, and
it has been shown \cite{haberweldon,yamadayaffe} that there is a well-defined
thermodynamic ensemble for any set of mutually commuting conserved charges.

Formally, we consider a system where the $\alpha_i$ fluctuate thermally about
some averages $\left<\alpha_i\right>$, and the energy fluctuates about an
average $\left<E\right>$. The partition function for this ensemble can be found
with various methods. For brevity we start with the Shannon/Jaynes
\cite{shannon,jaynes1} entropy
\begin{equation}\label{jaynes}
  \begin{split}
  S = -\sum_\sigma\Bigl[ \pi_\sigma \ln(\pi_\sigma)
  -\beta\Bigl(& \pi_\sigma H -\left<E\right>\\
  &-\sum_i\mu_i N\bigl(\pi_\sigma \alpha_i-\left<\alpha_i\right>\bigr)\Bigl)
  \Bigr] \; ,
  \end{split}
\end{equation}
where we have set $k_\mathrm{B}=1$,
$\pi_\sigma$ is the probability of finding the
system in a state labelled $\sigma$, $\beta$ and $\mu_i$ are Lagrange multipliers
enforcing the thermal averages, $N$ is the number of particles in the
system (the factor of $N$ is included here so that both $\mu_i$ and $\alpha_i$
can be intensive quantities),
and the summation should be interpreted schematically. Unless
otherwise noted we will work in units where the particle volume $\ell^3=1$.
To determine
the partition function we maximize Eq.\ \eqref{jaynes} with respect to $\pi_\sigma$.
This gives, up to some normalization constant $\Zcal$,
\begin{equation}
  \pi_\sigma = \frac{1}{\Zcal} e^{-\beta(H-\sum_i\mu_i N\alpha_i)} \; ,
\end{equation}
and fixing the normalization $\sum_\sigma \pi_\sigma=1$ gives
\begin{equation}\label{partfn}
  \Zcal = \sum_\sigma e^{-\beta(H-\sum_i\mu_i N\alpha_i)} \; .
\end{equation}
We see that $\beta=1/T$, the usual inverse temperature, and $\mu_i$ are
generalized chemical potentials conjugate to the ``charges'' $\alpha_i$
\cite{yamadayaffe} that determine the building block attributes. To distinguish
$\mu_i$ from the ordinary chemical potential, and since they act as sources for
changing the ``elemental'' building blocks of the system, we refer to them as
``alchemical'' potentials. We define the thermodynamic potential for this
ensemble as $\Zcal \equiv e^{-\beta\phi}$, which gives
\begin{equation}
  \left<\alpha_i\right>=-\frac{1}{N}
  \left(\frac{\partial\phi}{\partial\mu_i}\right)_{N,\eta,T,\mu_{j\neq i}} \; ,
\end{equation}
where $\eta$ is the packing fraction or density.
This computes
how the system responds to a change in alchemical potential, and
in the thermodynamic limit (hereafter we will be concerned about the
thermodynamic limit so we will drop the $\left<{}\right>$
notation) establishes a constitutive relation
$\alpha_i(\eta,T,\{\mu_j\})$. It is convenient to make a Legendre transformation $F=\phi+\sum_i\mu_i
N\alpha_i$, and compute the constitutive relation
$\mu_i(\eta,\beta,\{\alpha_j\})$ using the expression
\begin{equation}\label{eos}
  \mu_i=\frac{1}{N}\left(\frac{\partial
  F}{\partial\alpha_i}\right)_{N,\eta,T,\alpha_{j\neq i}} \; .
\end{equation}
For notational simplicity, especially in cases where we consider a single
$\alpha_i$, it will sometimes be convenient to drop the subscripts on
$\alpha$ and $\mu$.

The constitutive relation $\mu(\alpha)$ quantifies the thermodynamic response of
a system to a change in the attributes of the constituent
particles. See \SI\ for a discussion of higher order thermodynamic
response functions.
If the alchemical potential $\mu>0$ at some state point $NVT\alpha$, then an
infinitesimal increase in the alchemical parameter $\alpha$ would increase the
free energy of the system. Conversely if $\mu<0$ then an infinitesimal increase
in $\alpha$ would decrease the free energy of the system.
This has two important implications.
(i) Locally optimal particle attributes $\alpha^*$ are determined by the roots
of the constitutive relation $\mu(\alpha^*)=0$ with positive slope. We show in
\SI\ that the locations of these roots are invariant under reparametrizations of
$\alpha$.
(ii) In hard particle systems, where the free energy simply measures the system
entropy, $\mu$ directly measures how the number of states available to a system
changes as a function of the particle shape, and so it can be used to
systematically determine which particle features are most likely to come into
contact, and provides explicit quantitative guidance on how to design shapes for
structures. We demonstrate both of these implications below.

In the next section, we explicitly compute $\mu$ in three example systems, and
interpret the meaning and implications of each computation. We compute the
constitutive relation $\mu_i(\eta,T,\{\alpha_j\})$ at $\eta,T,\{\alpha_j\}$
numerically using Eq.\ \eqref{eos} with the Bennett acceptance ratio method
\cite{bennettacpr}. Using this method, we compute $\mu$ at some $\{\alpha_j\}$
by equilibrating several independent samples at nearby values $\alpha_j+\nu
h_j$, where $\nu$ are constants chosen for an appropriate finite differencing
scheme, and $h_j$ are finite differences. For a full description of the
computation, see \SI. To determine valence for anisotropic particles, we use the
potential of mean force and torque (PMFT), as described in Ref.\
\cite{entint}.

In addition to constitutive relations between thermodynamic quantities (\ie\
first order derivatives of the free energy), physical systems are also
frequently characterized by higher free energy derivatives: susceptibilities and
moduli (see, \eg, Refs. \cite{LLv5,LLv7}). We define
the alchemical modulus $M_\alpha$ and susceptibility $\chi_\alpha$ as
\begin{equation}\label{M_alpha}
  M_\alpha\equiv
  \left(\frac{\partial \mu}{\partial \alpha}\right)_{N,\eta,T} \; , \quad
  \chi_\alpha\equiv\left(\frac{\partial \alpha}{\partial \mu}\right)_{N,\eta,T}
  \; .
\end{equation}
The extension to systems with several alchemical parameters is straightforward.
We note that, like standard moduli (\eg\ bulk, shear, Young's), $M_\alpha$ is a
stress-strain relationship \cite{LLv7}, but the strain is in alchemical space rather
than real space. Accordingly, alchemical modulus $M_\alpha$ Eq.\ \eqref{M_alpha} at $\alpha^*$
measures how sensitive the system is to deviations from the ideal particle
properties. Similarly, like standard susceptibilities (\eg\ compressibility)
\cite{LLv7}, $\chi_\alpha$ is a strain-stress relationship. Physically, \eg, by
the fluctuation-dissipation theorem (see, \eg, Ref.\ \cite{LLv5}) $\chi_\alpha$ determines how
quickly a system of, say, fluctuating shape relaxes when particles are perturbed from
their equilibrium attributes.

\section{Numerical Results and Discussion}
We use our digital alchemy methodology to optimize building blocks for
self-assembly in three different case studies. The first two involve
entropy-driven systems, which are among the most conceptually difficult in which
to connect macroscopic and microscopic system properties because the macroscopic
behaviors are intrinsically
collective.\cite{kamien,epp,entint,escoengent,ordviaent} In the third study, we
investigate an oscillating pair potential, which was recently shown \cite{opp}
to self-assemble a one-component icosahedral quasicrystal, one of the most
complex crystal structures known. In each case, the details of the specific model are
included in the discussion below.  Details and extended discussion of the methods used in each case may be found in the SI.

\begin{figure*}[t]
  \begin{center}
    \includegraphics[width=16cm]{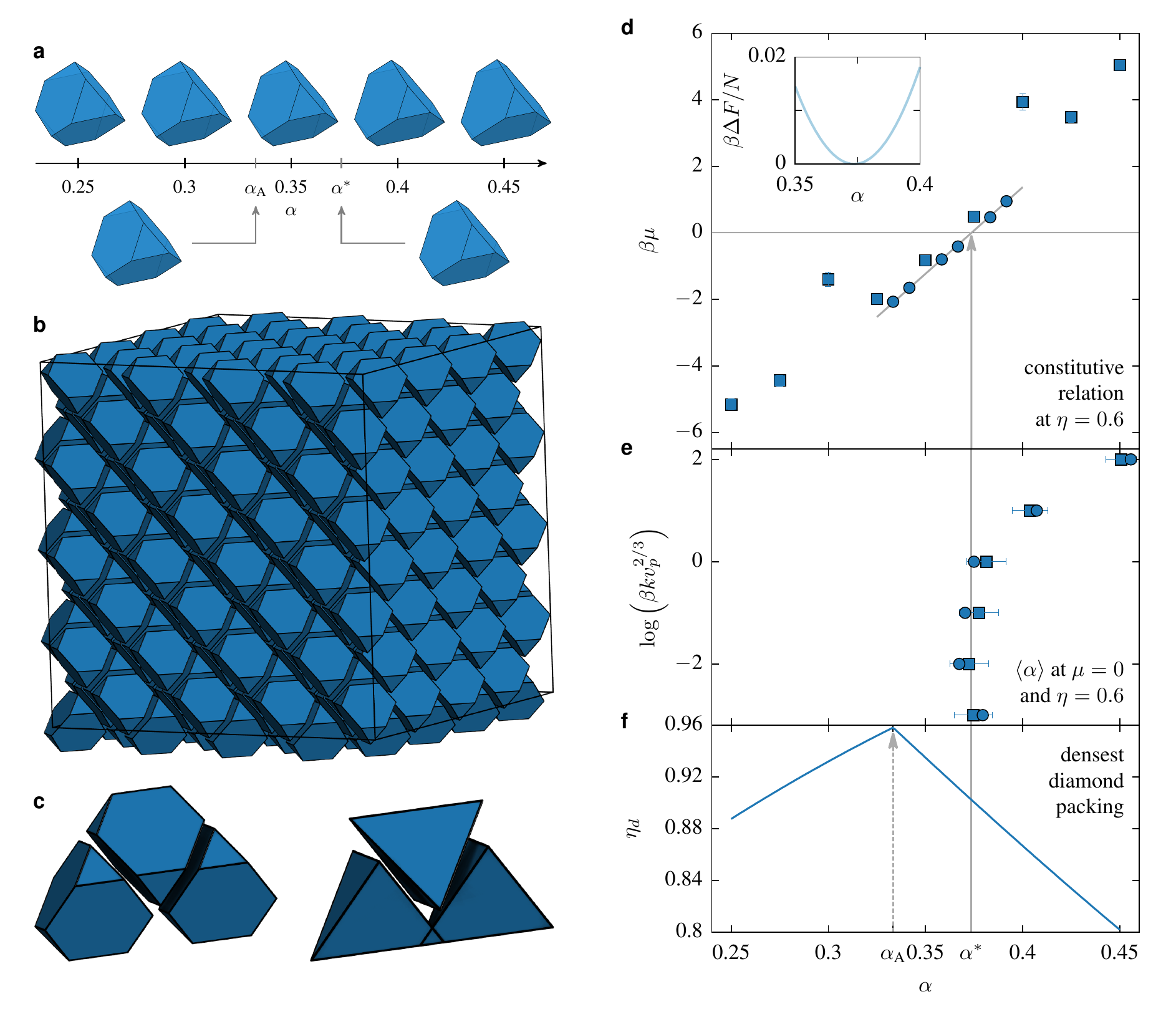}
  \end{center}
  \caption{Truncated tetrahedra at a range of truncations $\alpha$ (a)
  self-assemble a diamond lattice (b) \cite{trunctet}. A search for maximal
  diamond packing density, $\eta_d$ (f) would suggest optimal
  assembly at $\alpha_\mathrm{A}=1/3$, the Archimedean truncated tetrahedron.
  We compute the
  constitutive relation (d) $\mu(\alpha)$ (Eq.\ \eqref{eos}) for hard
  truncated tetrahedra at $\eta=0.6$. Squares (\protect\tikz
  \protect\draw[draw=black,fill=gcb2] (-0.08cm,-0.08cm) -- (0.08cm,-0.08cm) --
  (0.08cm,0.08cm) -- (-0.08cm,0.08cm) -- cycle;) are results for systems with
  $216$ particles, circles (\protect\tikz \protect\draw[draw=black,fill=gcb2]
  (0,0) circle (0.08cm);) for systems with $1000$ particles; where visible,
  error bars are one standard deviation. The alchemical potential vanishes when the
  truncation is optimal for self-assembling diamond at this density when the
  truncation is approximately $\alpha^*\approx0.37$.
  In the inset plot we reconstruct the free energy curve in the
  vicinity of the minimum. The increase in anisotropy $\alpha^*$ above the
  geometric prediction $\alpha_\mathrm{A}$ arises because particles need to
  increase anisotropy to preserve tetrahedral valence at lower packing
  fractions, but if particles are too anisotropic, the simultaneous coordination
  of neighboring particles is sterically prohibited (c, see also Fig.\
  \ref{tetvalence}). We demonstrate this
  design rule (e; see also \SI\ Movie) by simulating tetrahedra with fluctuating
  shape at $\mu=0$ in an Einstein crystal with spring constant $k$ at packing
  fraction $\eta=0.6$ and allowing the particles to find their optimal shape.
  The plot (e) shows that at low $k$ the average truncation
  $\left<\alpha\right>$ is consistent with $\alpha^*$ ($N=216$ squares \protect\tikz
  \protect\draw[draw=black,fill=gcb2] (-0.08cm,-0.08cm) -- (0.08cm,-0.08cm) --
  (0.08cm,0.08cm) -- (-0.08cm,0.08cm) -- cycle;; $N=1000$ circles
  \protect\tikz \protect\draw[draw=black,fill=gcb2]
  (0,0) circle (0.08cm);).
  \label{eos_tt}}
\end{figure*}

\begin{figure}[t]
  \begin{center}
    \includegraphics[height=18cm]{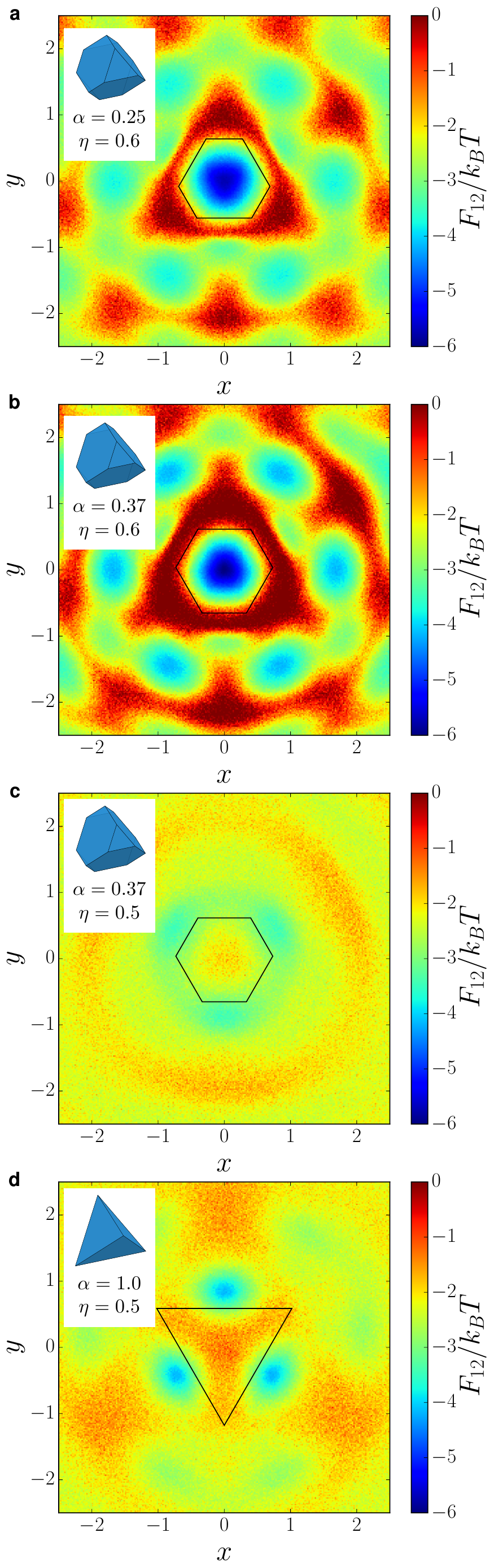}
  \end{center}
  \caption{Emergent valence encoded in the PMFT for truncated tetrahedra for a
  crystal at density $\eta=0.6$ (a: $\alpha=0.25$, b:
  $\alpha=\alpha^*\approx0.37$), and a fluid at density $\eta=0.5$ (c:
  $\alpha=\alpha^*\approx0.37$, d: $\alpha=1.0$). In the crystal we see that
  the particle at the optimal truncation $\alpha^*$ (b) shows greater
  specificity of tetrahedral valence than at lower $\alpha$ (a), as expected.
  However, at fluid densities, we see that if the particle is too tetrahedral
  (d), the second neighbor shell is by $\pi/6$ compared with lower truncations
  (c), and is incommensurate with the diamond lattice.
  \label{tetvalence}
  }
\end{figure}

\begin{figure}[t]
  \begin{center}
    \includegraphics[width=8.7cm]{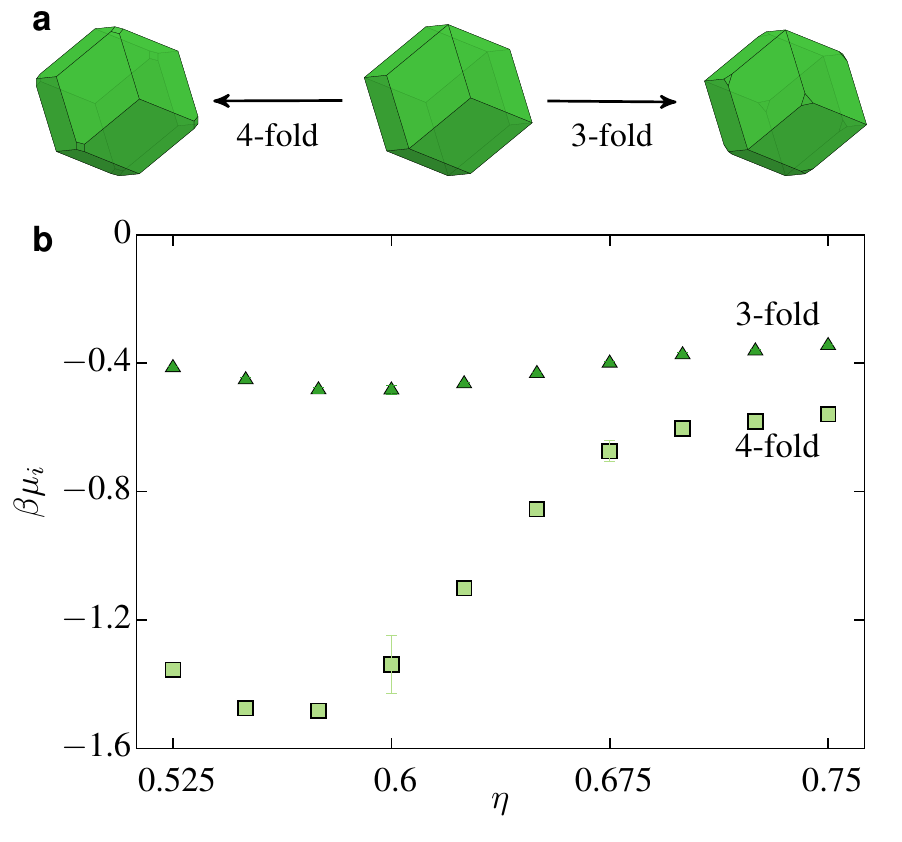}
  \end{center}
  \caption{
  Rhombic dodecahedra have both four-fold and three-fold vertices (a). We
  determine the relative sensitivity to the truncation of each type of vertex by
  computing the constitutive relation $\mu(\eta)$ (b) according to the
  (exaggerated) truncations shown in (a).
  We plot alchemical potentials for 4-fold truncations
  ($\mu_4$, \protect\tikz \protect\draw[draw=black,fill=gcb3] (-0.08cm,-0.08cm) --
  (0.08cm,-0.08cm) -- (0.08cm,0.08cm) -- (-0.08cm,0.08cm) -- cycle;
  squares)
  and 3-fold truncations
  ($\mu_3$, \protect\tikz \protect\draw[draw=black,fill=gcb4] (0,0.08cm) --
  (-0.069282cm,-0.04cm) -- (0.069282cm,-0.04cm) -- cycle;
  triangles) at various
  densities for which the system self-assembles an fcc lattice.
  We observed $\mu_4<\mu_3<0$ at all densities, indicating that both vertex
  truncations improve assembly of the target crystal, but four-fold vertex
  truncations provide greater improvement.
  \label{trunc_rdd}
  }
\end{figure}

\subsection{Truncated Tetrahedra}
We simulated a one-parameter family of truncated tetrahedra at moderate
truncations known to self-assemble diamond lattices \cite{trunctet}. We
parametrized the truncation between $\alpha=0$ (a tetrahedron maximally truncated
so that it is an octahedron) and $1$ (an untruncated, regular tetrahedron). With
this parametrization (we discuss reparameterization invariance of our
results in \SI) particles self-assembled diamond at a packing fraction of
$\eta=0.6$ between truncations of $0.25$, and $0.475$ (see
Fig.\ \ref{eos_tt}a). For reference, the Archimedean truncated tetrahedron
\cite{trunctet} has a truncation of $\tfrac13$. We performed standard Monte
Carlo simulations (\eg\ Ref.\ \cite{amirnature}) of systems of $N=216$ and $1000$
particles at fixed volume. Polyhedra overlaps were checked using the GJK
algorithm \cite{gjk}.

For the truncated tetrahedra, we computed the constitutive relation between
vertex truncation $\alpha$ and its conjugate alchemical potential $\mu$.  We
first computed $\mu$ in small systems of $N=216$ particles, and found preliminary
evidence for vanishing alchemical potential (here, a free energy minimum) for
$0.35<\alpha^*<0.4$, (Fig.\ \ref{eos_tt}d, squares). To obtain higher precision,
and to test for finite size effects, we simulated systems
of $N=1000$ particles in the region surrounding the putative free energy minimum
(Fig.\ \ref{eos_tt}d, circles). From these alchemical potential computations we
extracted the free energy of the system as a function of shape in the vicinity
of the minimum (Fig.\ \ref{eos_tt}d, inset), which we estimated by performing a
weighted least squares fit to
\begin{equation}
  \beta\mu = \beta M_\alpha (\alpha-\alpha^*) \; ,
\end{equation}
from which we find the free energy minimum is at
\begin{equation}
  \alpha^* = 0.3736 \pm 0.0001 \; ,
\end{equation}
and the alchemical modulus $M_\alpha$ is
\begin{equation}
  \beta M_\alpha(\alpha=\alpha^*,\eta=0.6)=52.0\pm0.3 \; .
\end{equation}
We also constructed diamond densest packings (Fig.\ \ref{eos_tt}f) for
truncated tetrahedra for all truncations (in increments of $0.001$) at
which self assembly into diamond lattices was reported in Ref.\ \cite{trunctet},
and find the curve has a maximum consistent with the Archimedean truncated
tetrahedron at $\alpha_\mathrm{A}=1/3$.

To directly examine the effects of shape modification on emergent valence
\cite{epp,entint}, we computed the PMFT for systems of $N=1000$ truncated
tetrahedra. For details of this computation, see Ref.\ \cite{entint}. We computed
the PMFT at a density of $\eta=0.6$ for a truncation of $\alpha=0.25$ (Fig.\
\ref{tetvalence}a), and a truncation of $\alpha^*$ (Fig.\ \ref{tetvalence}b).
The results for the first neighbor shell show particles have stronger
tetrahedral valence at $\alpha^*$ than at $\alpha=0.25$, which originates from
the relatively larger hexagonal faces acting as stronger entropic patches
\cite{epp}.  However, we see that at a fluid density of $\eta=0.5$, in the
second neighbor shell when particles have the optimal truncation $\alpha^*$
(cyan spots in Fig.\ \ref{tetvalence}c), the next-to-nearest neighbors sit in an
alternating arrangement, whereas the next-to-nearest neighbors for perfect
tetrahedra (blue spots in Fig.\ \ref{tetvalence}d) are rotated by $\pi/6$. This
indicates a non-alternating arrangement that coincides with polytetrahedral
motifs not commensurate with the diamond lattice, which arises from steric
constraints depicted in Fig.\ \ref{eos_tt}c. To directly confirm this result, we
performed simulations in an $NVT\mu$ ensemble (\ie\ both thermostated and
``alchemostated'') to determine $\alpha(\mu)$ at $\mu=0$ for $N=216$ and $1000$
truncated tetrahedra with fluctuating shape in a diamond Einstein crystal. We
initialized the system at low packing fraction $\eta=0.2$ with fully truncated
(\ie\ octahedral, $\alpha=0$) particles, and slowly compressed the system to
the target packing fraction of $\eta=0.6$, after which we relaxed the spring
constant. We observed that the process drove the particles to spontaneously
adopt a truncation consistent with our alchemical potential calculations at
fixed shape. See Fig.\ \ref{eos_tt}e, and \SI\ movie.

Our computation of the constitutive relation $\mu(\alpha)$ for truncated
tetrahedra that form a diamond lattice reveals several findings.
(i) By determining that $\mu(\alpha)$ has a root at $\alpha^*\approx0.37$ we have
demonstrated that it is possible to find a thermodynamically optimal shape, among
a given family, for self-assembling the diamond lattice.
(ii) Our criterion of $\mu(\alpha^*)=0$ is both parameter-free and independent of
system kinetics, which are highly dependent on simulation methods. Nevertheless,
we find rough agreement between the thermodynamic computation
of the alchemical potential and a measurement of the lower critical packing fraction
$\eta_c$ reported in Ref.\ \cite{trunctet}.
(iii) The fact that the optimal particle shape ($\alpha^*\approx0.37$) for
diamond assembly at $\eta=0.6$ is more tetrahedral than the optimal shape for
diamond packing ($\alpha_\mathrm{A}=1/3$), but not perfectly tetrahedral
($\alpha=1$), arises from a competition between two effects. Particles must have
tetrahedral valence to form the diamond lattice, but in the diamond lattice,
particles are arranged in an alternating motif (Fig.\ \ref{eos_tt}c,d). Shape
entropy considerations \cite{entint} suggest that as the system density is
lowered, particles must have larger entropic patches \cite{epp} to maintain
their emergent valence, as shown in Fig.\ \ref{tetvalence}. However, as
illustrated in Fig.\ \ref{eos_tt}d, if the particles are too tetrahedral, then
the alternating diamond motif leads to overlapping next-to-nearest neighbor
particles, as shown in Fig.\ \ref{tetvalence}. Hence, the optimal truncation of
a tetrahedron to self-assemble diamond is more tetrahedral than packing would
dictate to preserve valence, but not too tetrahedral to prevent particles from
having alternating valence.
(iv) We computed the alchemical modulus $M_\alpha$ for truncated tetrahedra at
$\eta=0.6$ and $\alpha=\alpha^*$. In future work it would be interesting to
determine how this modulus varies across system density in this system and
differs between systems/structures, or relates to effects of polydispersity, and
how it behaves at phase boundaries.
(v) The entropic assembly of anisotropic hard shapes is driven by emergent
valence\cite{epp,entint}, manifesting in directional entropic
forces\cite{trunctet}. A defining feature of emergent behaviors is that their
origin is difficult to trace to microscopic attributes of the system
constituents.\cite{moreisdifferent} Here, we explicitly demonstrate the general
principle that it is possible to optimize building block attributes, by which we
systematically control emergent valence, in order to optimally assemble a target
structure. Moreover, our results suggest a general design rule for entropic
valence: that as system density decreases, entropic patch size \cite{epp} must
increase to optimally assemble a dense packing phase. This design rule is
supported by another recent result \cite{sphinx} where it was found that for
several families of dimpled particles, the peak in packing density occurs at an
entropic patch size that is below the critical size for the \emph{onset} of
entropic assembly at low density. This is particularly strong evidence for the
design rule proposed here because the optimal patch size cannot be smaller than
the patch size at onset.
(vi) In practice, the synthesis of anisotropic colloidal particles is
often driven by a growth process that yields particles in a family of shapes.
Here we have shown, in an example family, how to optimally choose when to
terminate that growth process to obtain particles for assembling a specific
target structure.

\subsection{Rhombic Dodecahedron}
To (i) understand how to contrast the relative importance of different
shape modifications of a given shape, and (ii) determine how this relative
importance depends on system density, we studied a two-parameter family of
truncations of rhombic dodecahedra
that leave them invariant under the spheric triangle group
$\Delta_{4,2,3}$.\cite{dfamilyp} The $\Delta_{4,2,3}$ invariant family of shapes
is constructed with three families of planes that make up the faces of a cube, a
rhombic dodecahedron, and an octahedron, all oriented to preserve the necessary
point group symmetry. The rhombic dodecahedron has two different types of
vertices: four-fold vertices where four planes come together, and three-fold
vertices where three planes come together. Moving the planes that make up the
faces of the cube towards the origin truncates the four-fold vertices, and
moving the planes that make up the faces of the octahedron truncates the
three-fold vertices. We performed simulations that examine the effects of each
type of truncation on a perfect rhombic dodecahedron. We parametrize the vertex
truncations so that when $\alpha_4=0$ (four-fold vertex truncation) and
$\alpha_3=0$ (three-fold vertex truncation) the particle is a perfect rhombic
dodecahedron. Maximal truncation $\alpha_4=1$ and $\alpha_3=0$ yields a
perfect cube, and $\alpha_4=0$ and $\alpha_3=1$ yields a perfect
octahedron.

We determined how systems of perfect rhombic dodecahedra
($\alpha_4=\alpha_3=0$) respond to infinitesimal changes in $\alpha_3$ and $\alpha_4$.
 We computed the alchemical potentials $\mu_4$ conjugate to $\alpha_4$
(four-fold vertex truncations) and $\mu_3$ conjugate to $\alpha_3$ (three-fold
vertex truncations) for systems of $N=256$ rhombic dodecahedra at a series of
packing densities $\eta$ between $0.525$ and $0.75$ in increments of $0.025$ at
$\alpha_4=\alpha_3=0$. As shown in Fig.\ \ref{trunc_rdd} we find negative
alchemical potentials for both 3-fold and 4-fold vertex truncations
($\mu_3,\mu_4<0$) at all
densities studied $0.525\le\eta\le0.75$, implying that both types of vertex
truncation \emph{reduce} the free energy of the system. Moreover, we find that
truncation of the four-fold vertices results in a greater reduction in free
energy than the three-fold truncation.

Our computation of the constitutive relations $\mu_i(\eta)$ for rhombic
dodecahedra explicitly demonstrates how our methods can determine the relative
importance of various shape features. Determining the most important shape
features to control is crucial for anisotropic particle synthesis techniques,
and here we have demonstrated a general method for solving this problem. In
addition to providing this general proof-of-principle, our results have several
specific implications.
(i)~At all densities studied, we observed $\mu_{4}<\mu_{3}<0$ indicating that
both types of vertex truncation improve the self-assembly of rhombic dodecahedra
into a face-centered cubic (fcc) lattice. Because vertex truncation at fixed
volume means the particles become slightly more spherical, our result 
suggests that the structure is further stabilized by particles
exchanging some vibrational degrees of freedom for rotational ones.
Moreover, (ii)~because $\mu_4<\mu_3$ it suggests that the four-fold vertex
truncations are more important in restricting the rotational motion than the
three-fold vertices. There are $8$ three-fold vertices and $6$ four-fold
vertices in a rhombic dodecahedron, but the centroid-to-vertex
distance for a four-fold vertex is $4/3$ the distance for a three-fold vertex.
We might suspect that if a vertex type sticks out further from the shape, or
is greater in number, it will provide a greater steric constraint on the
microstates available to the system. Our result that $\mu_4<\mu_3$ suggests
that for the rhombic dodecahedron in an fcc lattice, the vertex distance is more
important than the number of vertices. It would be interesting to investigate
whether this design rule holds for other shapes, or is specific to rhombic
dodecahedra.
(iii)~Because the slopes of both $\mu(\eta)$ curves are positive for
$\eta\gtrsim0.6$, it suggests that particles give up rotational entropy faster
than translational entropy as the system density increases. We note that the
distinction between four-fold and three-fold vertices becomes smaller at larger
packing fractions, which suggests, surprisingly, that as the particles
increasingly lose rotational entropy the distinction between \textit{how} they
lose it becomes \textit{less} important. It would be interesting to see if this
result holds more generally in other systems.

\begin{figure}[t]
  \vspace{-0.5in}
  \begin{center}
    \includegraphics[width=6.8cm]{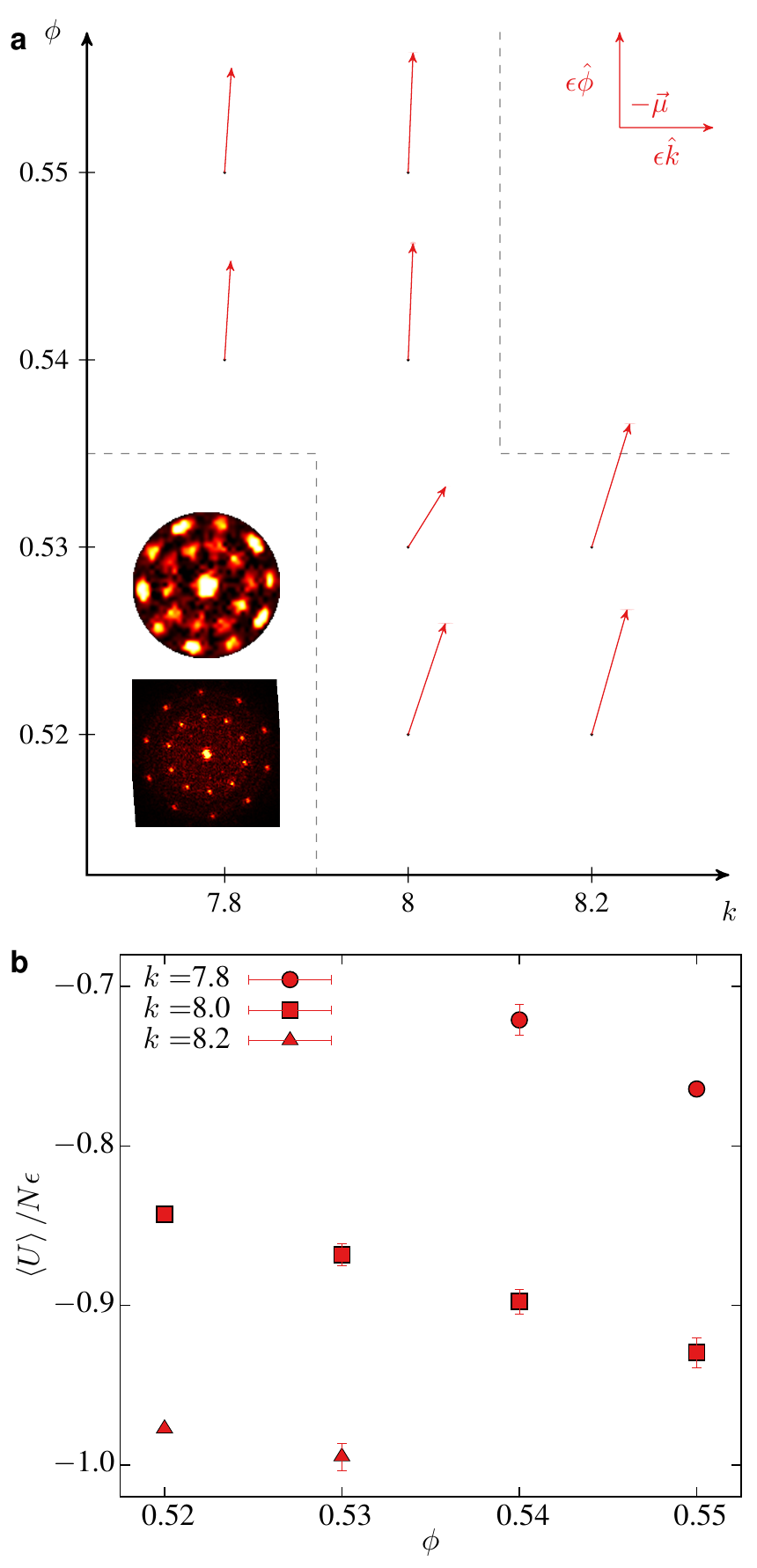}
  \end{center}
  \caption{
  Panel a: Alchemical potential ($-\vec{\mu}$) for the wavenumber $k$, and phase
  $\phi$ parameters systems of $N=4096$ particles interacting \textit{via} a
  three-well oscillating pair potential over a range of parameters that
  self-assemble an icosahedral quasicrystal.\cite{opp} The pair potential
  self-assembles icosahedral quasicrystals of three different densities. Here,
  we examine the region of parameter space that self-assembles the intermediate
  density quasicrystal (dashed lines indicate the phase boundaries we observed
  for the self-assembly of the intermediate density quasicrystal).
  Surprisingly, we find that, over the range of parameters we studied, in order
  to thermodynamically improve the assembly of the intermediate density
  quasicrystal, we are driven toward the region of parameter space that is
  dominated by the self-assembly of the high-density quasicrystalline phase.
  This suggests that the optimal choice of parameters to stabilize the
  intermediate density quasicrystal is buried in a region that will
  spontaneously self-assemble the high density phase instead, and suggests that
  the intermediate density phase will be difficult to stabilize in practice. The
  insets show the bond order diagram and diffraction pattern from a simulation
  snapshot of a $4096$ particle system at $k=8$, $\phi=0.53$.
  Panel b: In the same system we computed the average potential energy per particle
  for different values of $k$ as a function of $\phi$. We see a clear decrease
  in $\left<U\right>$ with increasing $k$ and $\phi$. This finding suggests that the
  decrease in free energy with increasing $k$ and $\phi$ shown in panel a can be
  attributed to enthalpic contributions from lower ground state energies.
  \label{oppfig}
  }
\end{figure}

\subsection{Oscillating Pair Potential}
To demonstrate that our alchemy approach is not limited to particle shapes, we
studied spherical nanoparticles (or point particles) interacting isotropically
using a truncated, intermediate range oscillating pair
potential studied in \cite{opp}, which is inspired by Friedel oscillations. It
can be written in the form
\begin{equation} \label{opp}
  U(r) =
  \frac{\epsilon}{r^{15}}+\frac{\epsilon}{r^3}\cos\left(k(r-1.25)-\phi\right) \; .
\end{equation}
This potential has been recently shown to self-assemble an icosahedral
quasicrystal
for $k_\mathrm{B}T=0.25$ for $0.78\lesssim k\lesssim0.82$ and
$0.52\lesssim \phi\lesssim 0.55$
\cite{opp}.
The potential is of particular interest due to the possibility of realizing it in
systems of nanoparticles or colloids decorated with appropriate ligands.
For these computations, we work in units with
$\epsilon=1$. 
We performed simulations of $N=4096$ particles
using HOOMD-Blue.\cite{hoomdblue} For full simulation details, see \SI.

We computed the alchemical potentials $\mu_k$ conjugate to $k$ (wavenumber) and
$\mu_\phi$ conjugate to $\phi$ (phase shift) for systems of $N=4096$ particles
interacting \via\ the oscillating pair potential in Eq.\ \eqref{opp}. We studied
the pair potential in the range of parameter space that was shown previously
\cite{opp} to self-assemble an intermediate density icosahedral quasicrystal. In
this phase, we find that within the entire parameter range over
which we were able to reliably nucleate the intermediate density quasicrystal,
both $\mu_k$ and $\mu_\phi$ are negative. We show this explicitly in Fig.\
\ref{oppfig}a where we form $\mu_k$ and $\mu_\phi$ into the vector $\vec{\mu}$.
We plot $-\vec{\mu}$, which shows the direction that decreases the free energy
at a given point in parameter space.

This result alone does not indicate whether this curious behavior is enthalpic
or entropic in origin.
To understand the origin of this decrease in free energy for increasing both $k$
and $\phi$, we computed the average potential energy at each state point, which
is plotted in Fig.\ \ref{oppfig}b. We see that at a given $k$, increasing $\phi$
decreases the system's potential energy, and that the potential energy is lower
at a given $\phi$ with increasing $k$, which is consistent with the alchemical
potential results shown in Fig.\ \ref{oppfig}a.
This suggests
that the effect we observe in Fig.\ \ref{oppfig}a is enthalpic in origin.

Surprisingly, our result that $\mu_k$ and $\mu_\phi$ are everywhere negative
suggests that there is not a choice of parameters for which $\vec{\mu}=0$ (\ie\
a local free energy minimum) in the parameter regime where the intermediate
density quasicrystal is the thermodynamically preferred phase. (For an example
of a simpler case where there is a local free energy minimum in a system with
isotropic interactions, see \SI.) Rather it suggests that, at least for systems
of $N=4096$ particles, the optimal parameter choice for self-assembling the
intermediate density quasicrystal lies along the boundary separating the
assembly of the intermediate density quasicrystal and the high density
quasicrystal, which is the thermodynamically preferred phase at higher values of
$k$ and $\phi$.\cite{opp}

A general take-away message of Ref.\ \cite{opp} is that controlling assembly in
one-component systems via isotropic interaction potentials involves two things.
It involves controlling the relative distances of potential energy
minima, which
determines preferred relative distances between particles. Note
that precise control over this procedure not straightforward, even at $T=0$. See
\SI\ for an explicit demonstration in a toy model system. However, it also
involves controlling the relative depth of the minima, which determines the
number of particles that sit at the preferred relative distances determined by
the minima locations. Here, we are able to directly compute the effects of
changes in potential control parameters on the system free energy and we find
that they can be detected. In \SI, we consider the pattern registration as
measured by comparing the locations of the potential minima with the radial
distribution function of the particles, and we find no discernible difference
across the range of parameters we considered. This suggests that our alchemical
potential methods are sensitive to system behavior that is not easily
discernible \via\ conventional analysis. We believe this might be of particular
value in systems such as the oscillating pair potential system where there is a
very rich bulk phase structure that depends sensitively on the choice of
potential parameters controlling particle valence \cite{opp}.

\begin{figure}[t]
  \begin{center}
  \begin{tabular}{ccc}
    \hline
    \hline
    & Atomic Matter & Colloidal Matter \\
    \hline
    \hline
    \makecell[c]{Anisotropy\\Dimensions} &
    Proton Number & Many \\
    \hline
    \makecell[c]{Anisotropy\\Dimension Types} &
    Discrete & \makecell[c]{Discrete,\\Continuous} \\
    \hline
    \makecell[c]{Valence\\Constraints} &
    \makecell[c]{Quantum Mechanics,\\Group Theory,\\Fermi Statistics} &
    Steric\\
    \hline
    \makecell[c]{Number of\\Stable Elements}
    & $\sim80$ & Infinite\\
    \hline
  \end{tabular}
  \end{center}
  \caption{Contrast between constraints on engineering materials with atomic
  elements and colloidal ``elements''. Colloidal elements have valence that can
  vary continuously in many different ways. The malleability of colloid valence
  means that constructing a ``periodic table'' for them is inherently difficult.
  However, we can exploit the malleability of colloids to directly probe how
  particle attributes affect structure.
  \label{colloidalmattertable}}
\end{figure}

\section{Concluding Remarks}\label{sec:disc}
We chose families of model systems to demonstrate the power of our methods
because of their structural complexity (the icosahedral quasicrystal), or
conceptual complexity (the emergent behavior of hard shapes); however our
methods can be generalized straightforwardly to systems of particles with other
interactions or shapes, as well as systems with enthalpic patches
\cite{patchy,glotzsolomon,pawarkretzchmar} or multiple particle species.
Furthermore, though our focus was on understanding macroscopic colloidal
behavior within a given region of phase space, our methods can be applied to the
crystallization of other types of matter, \eg\ polymers, and the study of phase
boundaries. One example where both are relevant is in the investigation of the
polymorphism \cite{polymorphism} or supramolecular isomerism in crystals of
small molecules, which is relevant for pharmaceutical applications
\cite{supramolecularisomerism}.

Here we focused on solving the problem of determining optimal building
block attributes for target structures among a range of building blocks,
from which we were able to extract design rules for emergent behavior. As
a result, most of our calculations were of the constitutive relation
$\mu(N,V,T,\alpha)$. However, for truncated tetrahedra we also considered (Fig.\
\ref{eos_tt}e and \SI\ Movie) the constitutive relation $\alpha(N,V,T,\mu)$ for
particles fixed to sit on a diamond lattice using a simple extension of Eq.\
\eqref{jaynes} (see \SI\ for details). All of the foregoing discussion
concerning interpretation of alchemical potentials, including the relation to
building block optimality, continues to hold, where any quantities computed in
extended ensembles are, by design, conditional on the externally imposed
criteria. Using extended ensembles, it is straightforward (see \SI\ for
details) to use our techniques for the discovery of building blocks for bulk
materials given a suitable choice of external design criteria. We leave a full
numerical investigation of this class of problems to future work.

Our method for determining optimal building blocks to self assemble
target structures
was based on the desire to make quantitative connections between building block
attributes and bulk behavior. To make our proof-of-principle demonstration
explicit, we ensured that the local minima we identified were
\textit{bona fide} global minima by computing exhaustively over relevant
building block attributes. Rather than compute exhaustively as we have here,
future investigations should reduce computational effort by employing global
optimization techniques. Indeed, work aimed at optimizing building
blocks for bulk attributes has employed genetic or evolutionary algorithms
\cite{camd,grartevo,dnageneticalgo,miskinsm,yinaml}, or gradient descent \cite{miskinpnas}.
Those approaches are complementary to the optimization part of the present work in
three ways:
(i)~Our approach provides a systematic, rigorous, first-principles method for
constructing probability distributions needed to apply the gradient descent
method proposed in Ref.\ \cite{miskinpnas}.
(ii)~Genetic and evolutionary algorithms are powerful techniques that use
external fitness criteria to perform non-local optimization. Our approach
supplements these non-local approaches by providing direct, precise measurement
of the physical response of a system to a local change in the attributes of
building blocks.
(iii)~The ability to probe local changes in building block attributes is also
important because, in addition to optimizing attributes, we would like to be
able to derive generalizable design rules that extend beyond specific systems of
interest. Here we showed an example of how to accomplish this using
digital alchemy by showing that dense packing arguments for anisotropic
shapes can be extended to lower density by increasing the size of entropic
patches.
We believe that a combination of the methods we present here with
existing techniques \cite{camd,grartevo,dnageneticalgo,miskinsm,miskinpnas} will
provide a powerful tool set for materials design.

Finally, our digital alchemy method shows how to phrase a generic class
of relationships between building block attributes and bulk behavior for
colloidal materials. For colloids, the fact that valence can vary in many ways
(sometimes continuously) along several different anisotropy dimensions
\cite{glotzsolomon,epp} means that it is not possible, even in principle, for a
periodic table of colloidal elements to be as succinct as the atomic periodic
table (see Fig.\ \ref{colloidalmattertable}). However, like the atomic periodic
table relates atomic valence to bulk behavior, we have shown that it is possible
to relate colloid valence to bulk behavior. Indeed, because colloid valence is
so malleable, we have shown that building block property--bulk behavior
relationships for colloids can be quantitative in a way that is not possible for
atoms. In effect, whereas quantum mechanics dictates that the atomic periodic
table is complete and succinct, but heuristic, the outlines for a periodic table
of colloidal elements suggested by this work are that it is complex and
many-dimensional, but also quantitative, and richly predictive.

\vspace{1em}\noindent%
\textit{
This document is an unedited Author's version of a Submitted Work that was
subsequently accepted for publication in ACS Nano, copyright (c) American
Chemical Society after peer review. To access the final edited and published
work see
\href{http://dx.doi.org/10.1021/acsnano.5b04181}{DOI:10.1021/acsnano.5b04181}.
}

\textit{
  We thank K.~Ahmed, J.~Anderson, D.~Beltr\'an-Villegas, J.~Crocker, E.~Eiser,
  D.~Frenkel, O.~Gang, L.~Isa, D.~Kofke, I.~Kretzschmar, R.~Newman, B.~Schultz,
  K.~Stebe, A.~Sweeney, and J.~Swift for helpful discussions and encouragement;
  C.~Phillips for providing simulation code for the Lennard-Jones-Gauss system;
  M.~Engel for providing simulation code for the polyhedra; J.~Antonaglia for a
  careful reading of an early version of the manuscript; J.~Dshemuchadse for
  helpful discussions and comments on the manuscript; P.~Damasceno for helpful
  discussions, encouragement, and assistance with structure identification; and
  H.~Jaeger for generously sharing a pre-publication version of Ref.\
  \cite{miskinpnas}. This material is based upon work supported by, or in part
  by, the U.S.\ Army Research Office under Grant Award No.\ W911NF-10-1-0518,
  the DOD/ASD(R\&E) under Award No.\ N00244-09-1-0062, and the Department of
  Energy under Grant No.\ DE-FG02-02ER46000. D.K.\ acknowledges funding by the
  FP7 Marie Curie Actions of the European Commission, Grant Agreement
  PIOF-GA-2011-302490 Actsa.
}

\renewcommand{\topfraction}{0.98}
\renewcommand{\dbltopfraction}{0.98}
\renewcommand{\textfraction}{0.02}
\renewcommand{\floatpagefraction}{0.9}
\renewcommand{\dblfloatpagefraction}{0.9}

\appendix
\section{Supplementary Theory and Methods}

\subsection{Moments of Inertia}
Our calculation of the alchemical potential that determines the constitutive
relation $\mu_i(\{\alpha_i\})$ depends on the moment of inertia tensor of the
system. To see this, we write the partition function for the ensemble with fixed
$\{\alpha_i\}$ as
\begin{equation} \label{Zisoschem}
  Z(\{\alpha_i\}) \propto \int [dp][dL][dq][dQ]e^{-\beta H} \; ,
\end{equation}
where $q$ are particle positions, $Q$ are particle orientations, $p$ are
conjugate momenta, $L$ are angular momenta, we have suppressed particle
indices, and for simplicity we are working in an ensemble with fixed volume and
number of particles. The following discussion is straightforward to extend to
other ensembles.

Starting with Eq.\ \eqref{Zisoschem}, we integrate over the momenta and angular
momenta, which gives (in three spatial dimensions)
\begin{equation} \label{Zreduce}
  \begin{split}
    Z(\{\alpha_i\}) \propto& \beta^{-3N} m^{3N/2}
    \left(\det(I_{\{\alpha_i\}})\right)^{N/2}\\
    &\int [dq][dQ]e^{-\beta U_{\{\alpha_i\}}(q,Q)} \; .
  \end{split}
\end{equation}
For simplicity, we will concern ourselves with changes in alchemical parameters
that leave the particle mass and volume invariant so that we are
isolating the effects of changes in shape only. Defining
\begin{equation} \label{Ztilde}
  e^{-\beta \tilde F} \equiv \int [dq][dQ]e^{-\beta U_{\{\alpha_i\}}(q,Q)} \; ,
\end{equation}
we have, up to irrelevant constants, the thermodynamic potential
\begin{equation} \label{potential}
  \beta F = -\frac{N}{2}\log\det\left(I_{\{\alpha_i\}}\right)
  +\beta \tilde F \; .
\end{equation}
We compute alchemical potentials by differentiating this expression with
respect to the $\alpha_i$. Even if the particle mass and volume are
fixed, the first term depends on the particle shape, and so we need to compute
the moment of inertia
tensor of our particles.

This term does not contribute to the computation of isotropic (spherical
particles), however it is important for polyhedral particles. Our computation
used four steps. (i) We compute the moments of inertia by identifying all of the
faces of the polyhedron. (ii) We do a fan decomposition of the faces into
triangles. (iii) We use the point at the origin with the triangulation of each
of the faces to decompose the polyhedron into a set of tetrahedra. (iv) We use
standard formulae to compute the inertia tensor of the
tetrahedron.\cite{tetinertia} We checked that our code was correct by using it
to compute moments of inertia for several known shapes. As an additional check,
all shapes we considered are invariant under triangle group symmetries. Via
Schur's lemma (see, \eg\ \cite{georgi}), their moments of inertia tensors must be
proportional to the identity matrix, and we checked that our code gave results
consistent with this up to machine precision.

\subsection{Numerical Evaluation of Alchemical Potential}
To evaluate the contribution of the configuration integral, $\tilde F$,
to the alchemical potential we use a variant of the Bennett acceptance ratio
method.\cite{bennettacpr}

We evaluate the expression
\begin{equation}\label{muexpr}
  \mu_i = \frac{1}{N} \frac{\partial F}{\partial\alpha_i}
\end{equation}
using finite differences. For brevity, we will give formulae for a single
$\alpha$; the extension to multiple $\alpha$ is straightforward. For some finite
$h$, we estimate
\begin{equation}\label{finitediff}
 \frac{\partial F}{\partial \alpha} \approx
  \frac{1}{h} \sum_\nu \gamma_\nu F(\alpha+\nu h)
\end{equation}
where, $\gamma_\nu$ and $\nu$ are appropriate constants for some finite
differencing scheme \cite{fornberg}.
To suppress numerical errors we used a symmetric four-point scheme for
calculations involving truncated tetrahedra, the oscillating pair potential, and
the 2D Lennard-Jones-Gauss system (see below); for rhombic dodecahedra we
used a one-sided four point scheme.

For each state point $\alpha$, we independently equilibrated several copies of
the target crystal lattice at each nearby $\alpha_\nu=\alpha+\nu h$. For the
$N=216$ ($h=4\times10^{-4}$) systems of truncated tetrahedra, the $N=4096$
oscillating pair potential systems ($h_k=3\times10^{-5}$, $h_\phi=10^{-5}$ or
$2\times10^{-5}$), and the $N=1024$ ($h=10^{-5}$) systems of Lennard-Jones-Gauss
particles we self-assembled crystals from the fluid; for the $N=1000$
($h=10^{-4}$) systems of truncated tetrahedra and the $N=256$ systems of
truncated rhombic dodecahedra ($h=2\times10^{-3}$) we constructed the crystal
directly. Without loss of generality, we labelled a particular $\nu$ as $\nu_0$.
For the hard truncated tetrahedra and rhombic dodecahedron systems, for all $\nu
\neq \nu_0$ we repeatedly sampled states from the equilibrium distribution using
standard Metropolis Monte Carlo techniques (see, \eg\ \cite{amirnature}), and
computed the probability, according to the Metropolis criterion
\cite{metropolis}, of accepting a
trial Monte Carlo move of
the state from the ensemble with $\nu$ to the ensemble
with $\nu_0$. Similarly,
we repeatedly sampled states from the equilibrium distribution of the system at
$\nu_0$, and computed the probability, according to the Metropolis criterion, of
making a trial Monte Carlo move of the state from ensemble $\nu_0$ to each of
ensembles at the other $\nu$. For the oscillating pair potential and
Lennard-Jones-Gauss systems, we repeatedly sampled configurations from
equilibrium $NVT$ molecular dynamics trajectories coupled to a Langevin
thermostat in HOOMD-Blue.\cite{hoomdblue} At each sample point we computed the
potential energy in the system at a given $\nu$ as well as what the potential
energy would be if the particles maintained all of their positions but
interacted with a potential $\nu'$. By recording these potential differences we
constructed the probability, according to the Metropolis criterion, of making a
Monte Carlo move from one value of $\alpha$ to another.

To show how this computes the alchemical potential, we note that combining
Eq.\ \eqref{muexpr} with Eq.\ \eqref{finitediff} and exponentiating we have
\begin{equation}
    e^{-hN\beta\mu} \approx \,
    e^{-\sum_\nu \alpha_\nu\beta F(\alpha+\nu h)}
    = \prod_\nu e^{-\gamma_\nu \beta F(\alpha+\nu h)} \; .
\end{equation}
Decomposing the thermodynamic potential into the kinetic and configuration
components using Eq.\ \eqref{potential} we have
\begin{equation}
    e^{-hN\beta\mu} \approx 
    \prod_\nu e^{\gamma_\nu [\frac{N}{2}\log\det I(\alpha+\nu h) -
    \beta \tilde F(\alpha+\nu h)]} \; .
\end{equation}
We then use detailed balance to write the configurational part of the free
energy $\tilde F$ at $\alpha+\nu h$ in terms of the value at $\alpha+\nu_0 h$ to
get
\begin{equation}
  \begin{split}
    e^{-hN\beta\mu} \approx &
    \prod_\nu e^{\gamma_\nu [\frac{N}{2}\log\det I(\alpha+\nu h) -
    \beta \tilde F(\alpha+\nu_0 h)]}
    \\&\quad
    \left(\frac{p(\alpha+\nu_0 h|\alpha+\nu h)}{p(\alpha+\nu h|\alpha+\nu_0
    h)}\right)^{\gamma_\nu} \; .
  \end{split}
\end{equation}
where $p$ are the relevant transition probabilities we compute with the Bennett
acceptance ratio method. We note that, since we are evaluating a first derivative of
$F$, $\sum_\nu \gamma_\nu=0$ so that
\begin{equation}
  \begin{split}
    e^{-hN\beta\mu} \approx &
    \prod_\nu e^{\gamma_\nu \frac{N}{2}\log\det I(\alpha+\nu h)}
    \left(\frac{p(\alpha+\nu_0 h|\alpha+\nu h)}{p(\alpha+\nu h|\alpha+\nu_0
    h)}\right)^{\gamma_\nu} \; .
  \end{split}
\end{equation}
Finally, taking the logarithm of both sizes we get that
\begin{equation}\label{mufinitediff}
  \begin{split}
    \beta\mu
    \approx & \frac{1}{h}\sum_\nu \frac{\gamma_\nu}{2}\log\det I(\alpha+\nu h)\\
    &-\frac{1}{Nh}\sum_\nu \gamma_\nu \log
    \frac{p(\alpha+\nu_0 h|\alpha+\nu h)}{p(\alpha+\nu h|\alpha+\nu_0 h)} \; .
  \end{split}
\end{equation}

\begin{figure}
  \begin{center}
  \resizebox{8.5cm}{!}{\includegraphics{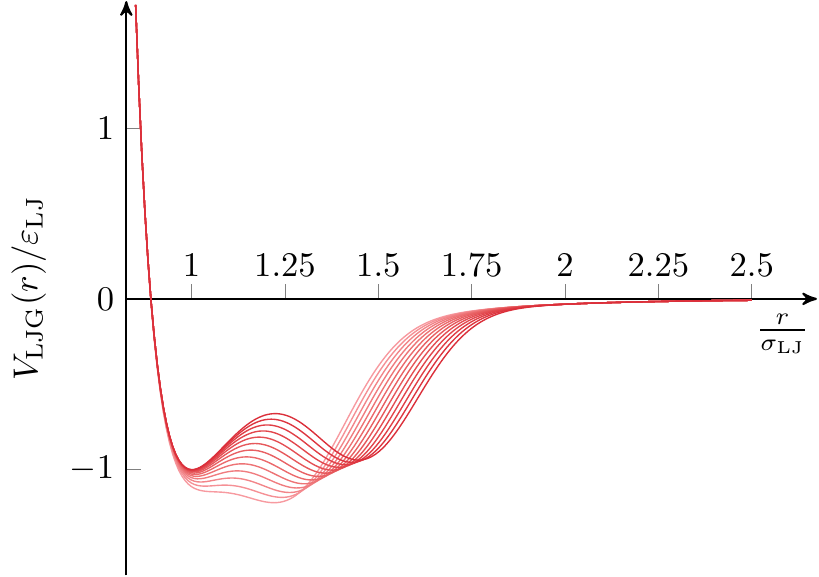}}
  \end{center}
  \caption{The functional form of the Lennard-Jones-Gauss potential (Eq.\
  \eqref{VLJG}) for $\varepsilon_\mathrm{G}/\varepsilon_\mathrm{LJ}=3/4$,
  $\sigma_\mathrm{G}/\sigma_\mathrm{LJ}=\sqrt{2}/10$, and
  $r_0/\sigma_\mathrm{LJ}$ in the range for a two dimensional system to form a
  square lattice.
  \label{ljg_func}
  }
\end{figure}

\begin{figure}
  \begin{center}
  \resizebox{8.0cm}{!}
  {
  \begin{tikzpicture}
    \node[draw=none,fill=none,anchor=north west] at (0,0)
    {\resizebox{8cm}{!}{\input{fig3_s}}};
    \node[circle,shading=ball,ball color=gcb6,minimum width=5mm] (oo) at (2cm,-1cm)
    {};
    \node[circle,shading=ball,ball color=gcb6,minimum width=5mm] (ox) at (3cm,-1cm)
    {};
    \node[circle,shading=ball,ball color=gcb6,minimum width=5mm] (oy) at (2cm,-2cm)
    {};
    \node[circle,shading=ball,ball color=gcb6,minimum width=5mm] (xy) at (3cm,-2cm)
    {};
  \end{tikzpicture}
  }
  \end{center}
  \caption{
  Constitutive relation $\mu(\alpha)$
  (\protect\tikz \protect\draw[draw=black,fill=gcb6] (-0.08cm,-0.08cm)
  -- (0.08cm,-0.08cm) -- (0.08cm,0.08cm) -- (-0.08cm,0.08cm) -- cycle;, top
  plot), where $\alpha=r_0/\sigma_\mathrm{LJ}$
  describes the potential well position, for a
  two-dimensional Lennard-Jones-Gauss system of $1024$ particles at
  $k_\text{B}T=0.1\varepsilon_\mathrm{LJ}$ (top).
  In this range of $\alpha$ the system forms a
  square lattice. The alchemical potential vanishes at $\alpha^*\approx 1.38$,
  indicating that that is the optimal well-location for self-assembling a square
  lattice. This accords with a calculation showing that the potential energy is
  minimized at this well position both $k_\text{B}T=0.1$ ({\protect\tikz
  \protect\draw[draw=black,fill=gcb6] (0,0.08cm) -- (-0.08cm,0) -- (0,-0.08cm)
  -- (0.08cm,0) --cycle;}, middle plot) and at $k_\text{B}T=0$ (solid line,
  middle plot), but is lower than the na\"ive
  geometric estimate of that comes from pattern registration of the potential
  minima to the
  ratio of nearest-neighbor to next to nearest-neighbor distances in the square
  lattice ($\sqrt{2}$, intercept in lower plot).
  \label{ljgalch}
  }
\end{figure}

Note that we did not attempt to determine general criteria for choosing optimal
finite differencing schemes, \ie\ the value of $h$ or the order of the method.
Two considerations that arise are that if $h$ is too large, the finite
differencing error is large, and a higher order method must be used. However, if
$h$ is too small, this affects the autocorrelation time of the transition
probabilities which, necessitates more extensive sampling. Figuring out how to
do this optimally for this type of computation is an interesting problem
for future work.

\begin{figure}
  \begin{center}
    \includegraphics[width=8.5cm]{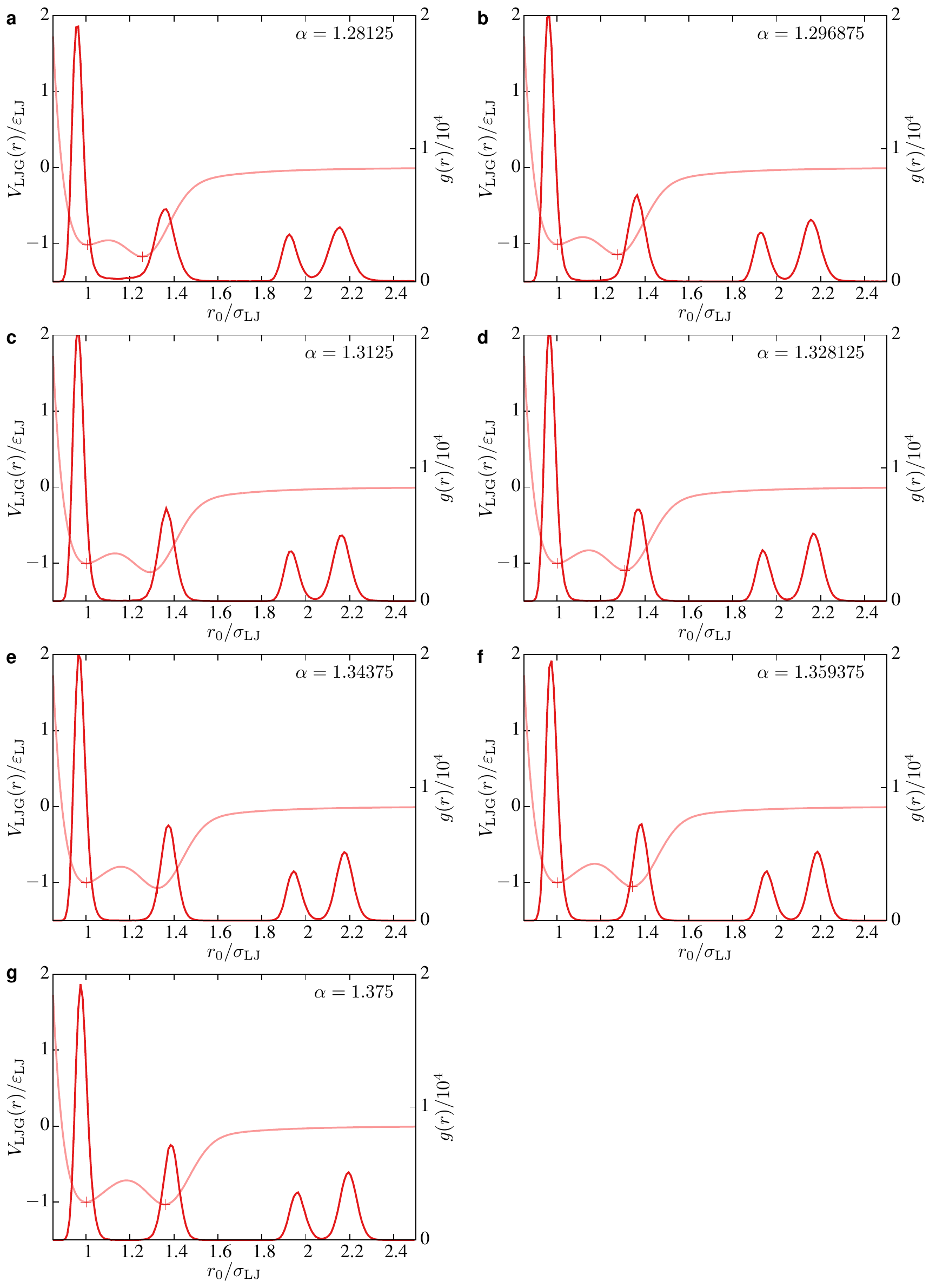}
  \end{center}
  \caption{We examine pattern registration of self-assembled square lattices in
  2D as shown through the radial distribution function ($g(r)$, darker curve)
  with the Lennard-Jones-Gauss pair potential (lighter curve, minima indicated
  with darker marks) that induces it to assemble, for various values of the
  parameter $\alpha=r_0/\sigma_\mathrm{LJ}$ (see Eq.\ \eqref{VLJG} in SI text).
  In these plots $r_0$ is less than the critical value of $\alpha^*$ where the
  alchemical potential $\mu$ vanishes.
  \label{r0less}
  }
\end{figure}

\begin{figure}
  \begin{center}
    \includegraphics[width=8.5cm]{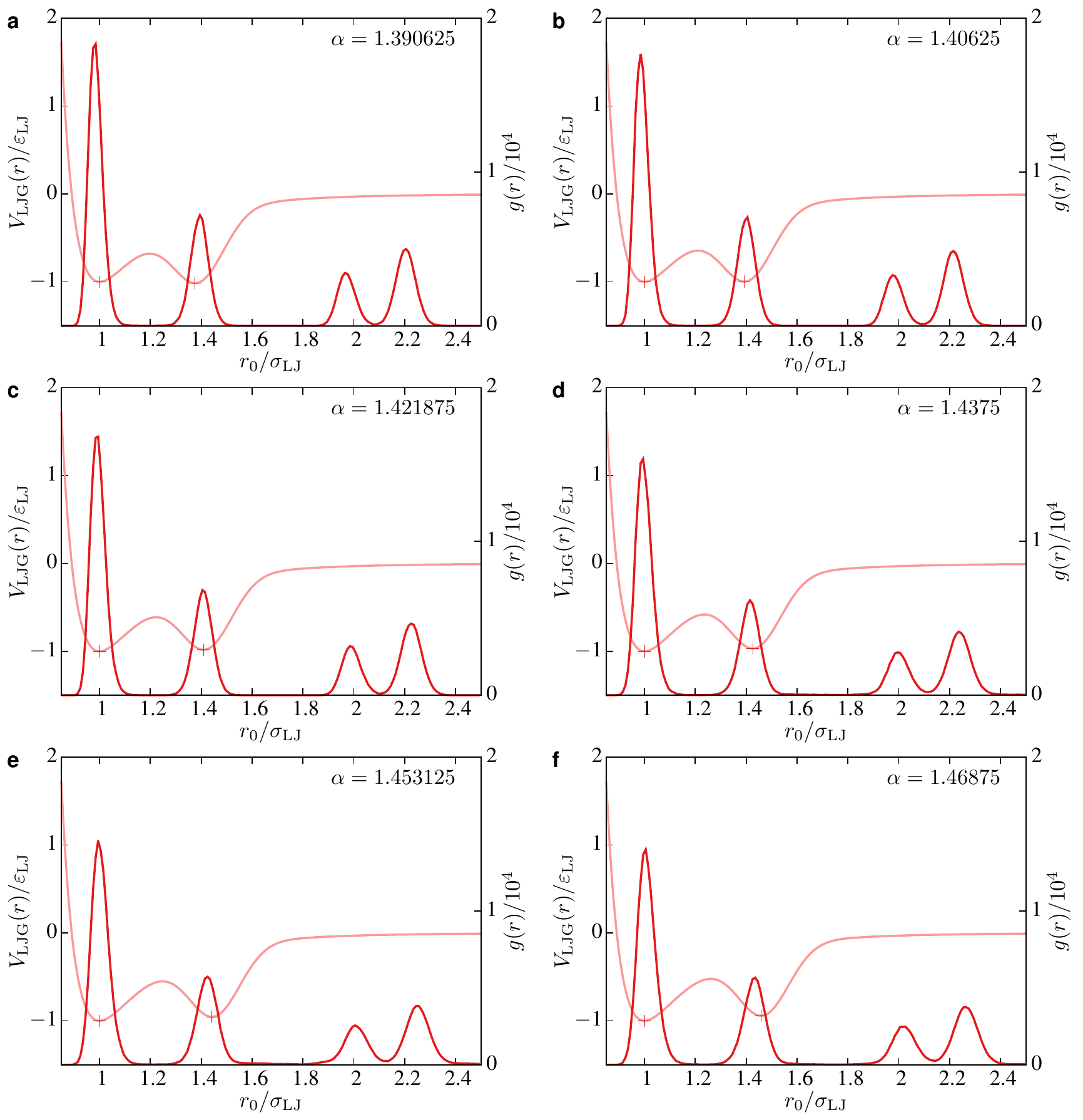}
  \end{center}
  \caption{We examine pattern registration of self-assembled square lattices in
  2D as shown through the radial distribution function ($g(r)$, darker curve)
  with the Lennard-Jones-Gauss pair potential (lighter curve, minima indicated
  with darker marks) that induces it to assemble, for various values of the
  parameter $\alpha=r_0/\sigma_\mathrm{LJ}$ (see Eq.\ \eqref{VLJG} in SI text).
  In these plots $r_0$ is greater than the critical value of $\alpha^*$ where
  the alchemical potential $\mu$ vanishes.
  \label{r0great}
  }
\end{figure}

\subsection{Error Estimation}
Numerical evaluation of the expression Eq.\ \eqref{mufinitediff} involves both
systematic and statistical error. Statistical error comes from the estimation of
the transition probabilities, and is given by
\begin{equation}\label{staterr}
  \begin{split}
  \delta(\beta\mu)_\text{stat} = \frac{1}{Nh}\sum_\nu \gamma_\nu \Bigl(&
    \frac{\delta p(\alpha+\nu_0 h|\alpha+\nu h)}
    {p(\alpha+\nu_0 h|\alpha+\nu h)}\\&
    + \frac{\delta p(\alpha+\nu h|\alpha+\nu_0 h)}
    {p(\alpha+\nu h|\alpha+\nu_0 h)}\Bigr)\; .
  \end{split}
\end{equation}

Systematic error can arise from the calculation of the moment of inertia tensor.
In practice, the moment of inertia tensor can be computed to machine precision,
so this contribution is negligible. Further systematic error can arise if
systems are equilibrated at slightly different densities. To see how this
arises, consider the differential of the free energy for our systems, which has
the form
\begin{equation}
  dF = \mu N d\alpha-PdV \; .
\end{equation}
where $N$ is the number of particles, and for simplicity we are considering a
single alchemical parameter $\alpha$ and working in ensembles with fixed volume.
To compute the alchemical potential $\mu$ we need to differentiate the free
energy at fixed volume. However, there can be
a variation in the system volume,
which in our systems is reflected in a change in the
packing fraction $\eta$, defined by
\begin{equation}
  V = \frac{N \ell^3}{\eta} \; ,
\end{equation}
where $\ell^3$ is the volume of a particle. Differentiating gives
\begin{equation}
  dV = - \frac{N \ell^3}{\eta^2} d\eta \; .
\end{equation}
So the differential for the free energy is given by
\begin{equation}
  dF = \mu N d\alpha+\frac{NP\ell^3}{\eta^2} d\eta \; .
\end{equation}
That means that to get a measurement of the alchemical potential, we need to
have
\begin{equation} \label{difflimit}
  \mu d\alpha \gg \frac{P\ell^3}{\eta^2} d\eta
\end{equation}
to safely control the error arising from changes in packing fraction. We did
this in two ways. For the systems of $N=1000$ truncated tetrahedra and $N=256$
rhombic dodecahedra we built perfect crystals at the desired packing fraction
and then thermally equilibrated them. \textit{E.g.}\ for the systems of $N=256$ rhombic
dodecahedra, this fixes the variation in packing fraction to $\lesssim 10^{-6}$, and so the
right hand side of Eq.\ \eqref{difflimit} is $\sim 10^{-5}$. On the left hand
side $d\alpha = 2\times 10^{-3}$, and
$\mu\sim10^{-1}$, which indicates that the
spread in packing fraction contributes a systematic error on the order of
$10\%$, which does not affect any of our conclusions. Systems of
$N=1000$ tetrahedra are similar, but in that case we also find agreement
between our alchemical potential computation and the computation with
fluctuating shape (\cf\ Fig.\ 2d,e, main text). For the systems of $N=216$
tetrahedra, we self-assembled crystals to within $1\%$ of the desired packing
fraction and then controlled for the errors in packing fraction statistically,
and again the results of that computation agree with the computation with
fluctuating shape.

For further confirmation, we note that our computation involves making trial
moves between different types of particles in the same system. That means that
when we compute the probability of going from $\alpha\to\alpha+\delta\nu h$, the
volume of the system is invariant. And, similarly, when we compute the
probability of going from $\alpha+\delta\nu h\to\alpha$, the volume is also
invariant. However, there is a small difference in density \emph{between} the
two systems. We tested controlling for this error in for four-fold vertex
truncations at a density of $\eta=0.625$ where we observed since we observed the
maximal sensitivity to packing fraction for that type of truncation at that
density. We controlled for the error by independently equilibrating systems at
packing fractions $\eta_i$ distributed near the desired packing fraction $\eta$.
We obtained a large number of independent samples of the transition
probabilities for the various $\eta_i$.  Given these transition probabilities,
we used regression (weighted by the errors in each of the transition
probabilities at $\eta_i$) to estimate the value of $p$ at $\eta$. We used the
error from the regression estimate as the statistical error in $\beta\mu$ in
Eq.\ \eqref{staterr}. To well-within the error bounds we found no difference between
the results obtained with densities randomly spread about the desired value, and
those set to the desired value to machine precision.

\subsection{Numerical Limits on Determining Optimal Building Blocks}
We note that in order to determine the roots of the constitutive
relation $\mu(\alpha)$ that determine optimal particle configurations to
arbitrary accuracy runs up against the limit Eq.\ \eqref{difflimit}. In practice we
note that reasonable computations on modern hardware allow us to bound optimal
shapes, for example, to differences in morphology that are imperceptible.

\subsection{Reparametrization and the Alchemical Constitutive Relation}
Our calculations of optimal shapes and interactions for target structures were
carried out by choosing a particular parametrization of the particle shape or
interaction potential. This choice was not unique. For simplicity, consider a
single parameter family of shapes or interaction potentials, and reparametrize
them by $\alpha=\alpha(\alpha')$. Under this reparametrization, the constitutive
relation becomes
\begin{equation}
  \mu'(\alpha') = \frac{\partial \alpha}{\partial \alpha'} \mu(\alpha) \; .
\end{equation}
From this form we note that if the reparametrization is
monotonic $d\alpha'/d\alpha>0$, then
$\sgn\left(\mu'(\alpha')\right)=\sgn\left(\mu(\alpha)\right)$, which means that
reparametrization will not change the direction of the change $\alpha$ that
corresponds to a decrease in free energy. Moreover, if $d\alpha'/d\alpha\neq0$,
any roots $\alpha^*$ of the original constitutive relation $\mu(\alpha)=0$ will
coincide with roots of the reparametrized constitutive relation $\mu'(\alpha')$
according to ${\alpha'}^*=\alpha'(\alpha^*)$, which means that particle shapes,
\eg, determined to be optimal from our alchemical potential calculations are
optimal regardless of the way in which particle shape is parametrized.

\section{Field Directed Alchemy: Design}
\subsection{Extended Ensembles}
In the main text, we are concerned with optimizing among known building blocks for
a target structure, so we began with Eq.\ (2) (main text) to derive and
interpret unbiased statistical ensembles. In order to design building blocks for
a structure for which we do not have a set of \textit{a priori} candidates, it is
necessary to modify Eq.\ (2) (main text) so that the statistical ensembles are
biased to form a structure with a desired property. To make this explicit, we
suppose there is some quantity $\Lambda$ that when evaluated on the design
structure takes the value $\left<\Lambda\right>$. The ensemble for this system
can be found by maximizing the entropy
\begin{equation}\label{jaynesengin}
  \begin{split}
  S = -\sum_\sigma\Bigl[ \pi_\sigma \ln(\pi_\sigma)
    -\beta\Bigl(& \pi_\sigma H -\left<E\right>\\
    &-\sum_i\mu_i N\bigl(\pi_\sigma \alpha_i-\left<\alpha_i\right>\bigr)\\
    &-\lambda\bigl(\pi_\sigma \Lambda-\left<\Lambda\right>)\Bigl) \; ,
  \Bigr]
  \end{split}
\end{equation}
with respect to $\pi_\sigma$, where $\lambda$ is a Lagrange multiplier. This
yields the extended partition function
\begin{equation}\label{Zengin}
  \Zcal = \sum_\sigma e^{-\beta(H-\sum_i\mu_i N\alpha_i-\lambda\Lambda)} \; .
\end{equation}
Note that $\lambda$, which is a Lagrange multiplier, determines the strength of
the coupling to the external field. We note that if $\lambda$ is positive
(negative), the system is driven toward particle configurations $\alpha_i$ that
favor increasing (decreasing) $\Lambda$, which allows one to design both
toward or away from a structural characteristic encoded in $\Lambda$. Moreover,
one could certainly use multiple structural characteristics with the aim of
arriving at building blocks suitable for reconfigurable structures, or the
suppression of some particular polymorph.

\subsection{Fluctuating Shape: Detailed Balance and Interpretation}
Here we show how to satisfy detailed balance in systems with fluctuating shape
aimed at directly solving the problem of particle design.
We begin with the generalized partition function from Eq.\ \eqref{Zengin}
\begin{equation}\label{Zengin_pL}
\Zcal =  \int d\alpha [dp] [dL] [dq] [dQ]
              e^{-\beta(H-\mu N\alpha-\lambda\Lambda)} \; ,
\end{equation}
where we have now explicitly included the measures for the integration over the
particle momenta $p$, angular momenta $L$, positions $q$, and orientations $Q$,
and for notational simplicity we use a single anisotropy parameter (the
generalized form is straightforward). Taking the Hamiltonian from Eq.\
(1) (main text), we perform the quadratic integrals over $p$ and $L$ to get
\begin{equation}\label{Zreducefull}
\Zcal \propto \int d\alpha [dq] [dQ] \left(\det(I_{\alpha})\right)^{N/2}
                   e^{-\beta(U_\alpha-\mu N\alpha-\lambda\Lambda)} \; ,
\end{equation}
up to irrelevant overall multiplicative constants. Detailed balance requires
that for a Markov chain Monte Carlo integration to converge to Eq.\
\eqref{Zreducefull} the ratio of the probability of making a move from a shape
$\alpha_1$ to a shape $\alpha_2$ $\Pi_{1\to 2}$ to the probability of the reverse
move $\Pi_{2\to 1}$ is equal to the ratio of probabilities of being in those
states
\begin{equation}\label{dbdef}
  \frac{\Pi_{2\to 1}}{\Pi_{1\to 2}} = \frac{\pi_1}{\pi_2} \; .
\end{equation}
Using Eq.\ \eqref{Zreducefull} we have
\begin{equation}\label{pratio}
  \frac{\Pi_{2\to 1}}{\Pi_{1\to 2}} =
  \frac{\left(\det(I_{\alpha_1})\right)^{N/2}
  e^{-\beta(U_{\alpha_1}-\mu N\alpha_1-\lambda\Lambda)}}
  {\left(\det(I_{\alpha_2})\right)^{N/2}
  e^{-\beta(U_{\alpha_2}-\mu N\alpha_2-\lambda\Lambda)}}
  \; .
\end{equation}
We performed simulations of truncated tetrahedra with fluctuating shape at zero
alchemical potential $\mu=0$, in an external field $\Lambda$ that forces the
particles to sit in an Einstein crystal with spring constant $k$ (measured in
units of $k_\mathrm{B}T/\ell^{2}$). Alchemical moves were performed at fixed
particle position and orientation, which allows us to simplify Eq.\
\eqref{pratio} to
\begin{equation}\label{pratiosimp}
  \frac{\Pi_{2\to 1}}{\Pi_{1\to 2}} =
  \frac{\left(\det(I_{\alpha_1})\right)^{N/2}}
  {\left(\det(I_{\alpha_2})\right)^{N/2}}
  e^{-\beta(U_{\alpha_1}-U_{\alpha_2})}
  \; .
\end{equation}
To satisfy detailed balance, we take the Metropolis
\cite{metropolis} criterion as 
\begin{equation}\label{metro}
  \Pi_{2\to 1} =
  \min\left(1,\frac{\left(\det(I_{\alpha_1})\right)^{N/2}}
  {\left(\det(I_{\alpha_2})\right)^{N/2}}
  e^{-\beta(U_{\alpha_1}-U_{\alpha_2})}\right)
  \; .
\end{equation}

\begin{figure}
  \begin{center}
    \includegraphics[width=8.5cm]{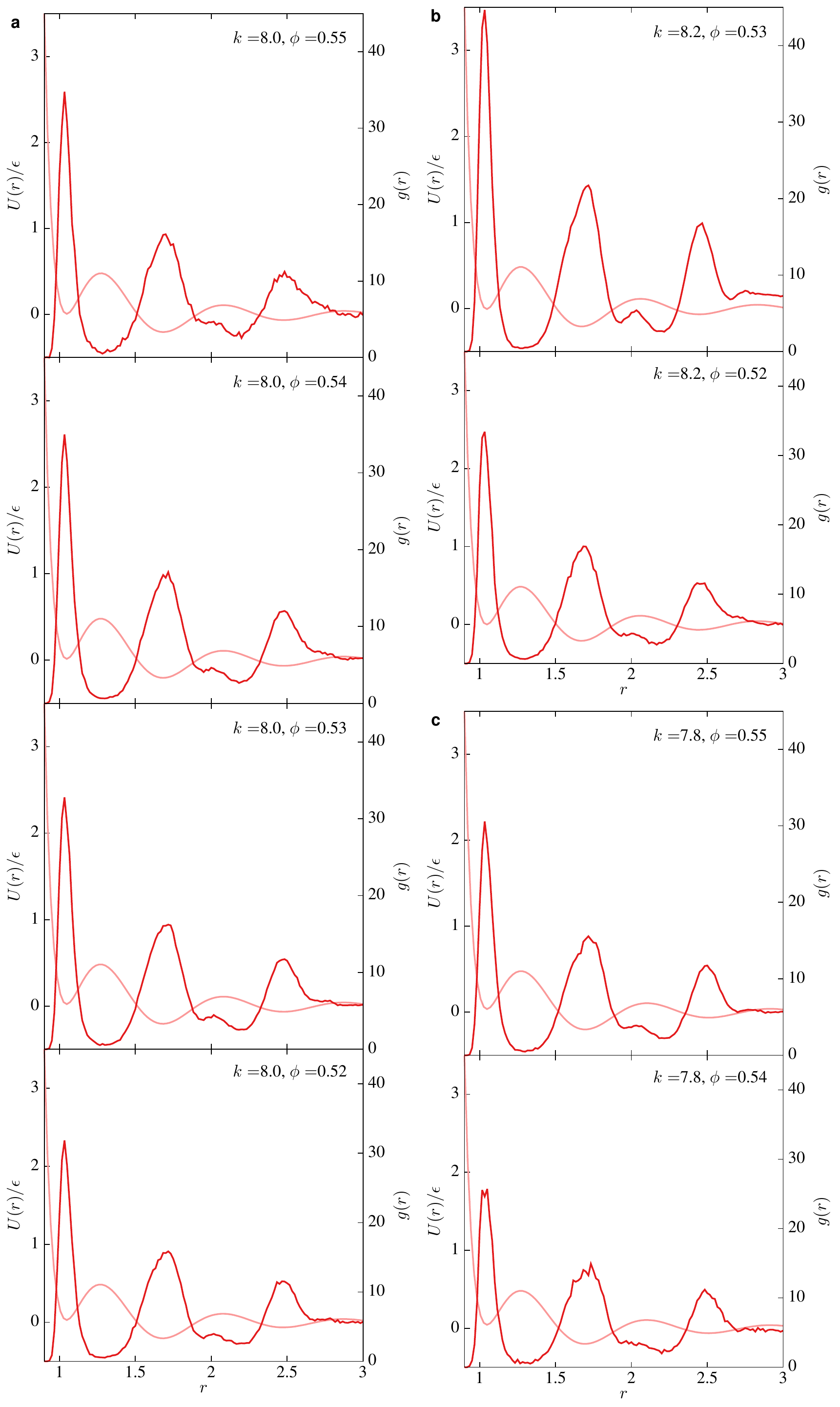}
  \end{center}
  \caption{
    For the oscillating pair potential at fixed $k$ (Panel a: $8.0$, b: $7.8$,
    c: $8.2$) we show the pattern registration as measured by coincidence of
    peaks in the radial distribution function for a snapshot of the icosahedral
    quasicrystal ($g(r)$, darker curve), with the pair interaction potential
    ($U(r,k,\phi)$, lighter curve). We see no discernible difference in pattern
    registration over this range of parameters.
    \label{fullstack}
  }
\end{figure}

In Fig.\ 1e (main text) we report $\alpha(\mu=0)$ for truncated tetrahedra at
a packing density of $\eta=0.6$ using an externally imposed field that puts the
particles in an Einstein diamond crystal with spring constant $k$. We show that
$\alpha(\mu=0)$ increases with $k$ (\ie\ particles become more tetrahedral for
large $k$). This result has two implications. (i) It shows that we can optimize
particle shape not only for the diamond structure, but that we can optimize
particle shape for a diamond structure at a fixed density with a stiffness that is
determined by the stiffness of the Einstein crystal we impose externally. This
suggests, more generally, that digital alchemy can optimize both
structures and properties of structures. (ii) It shows that one effect of making
particles more tetrahedral is that they form a diamond lattice that is more
stiff at fixed density. At large spring constants we observed $\alpha\approx0.5$,
however, we failed to observe the spontaneous assembly of diamond lattices in
our simulations at truncations that were this small. Moreover, our alchemical
potential calculations show that as the vertex truncation of tetrahedra
decreases past the optimal value, particles lose entropy in the diamond lattice.
These two results together show that though it is possible to tune the stiffness
of the crystal, which might be surprising because entropy is the only governing
property in these systems, the range over which additional properties can be
tuned while still ensuring spontaneous self-assembly is limited by kinetic
factors.

\begin{figure}
  \begin{center}
    \includegraphics[width=8.5cm]{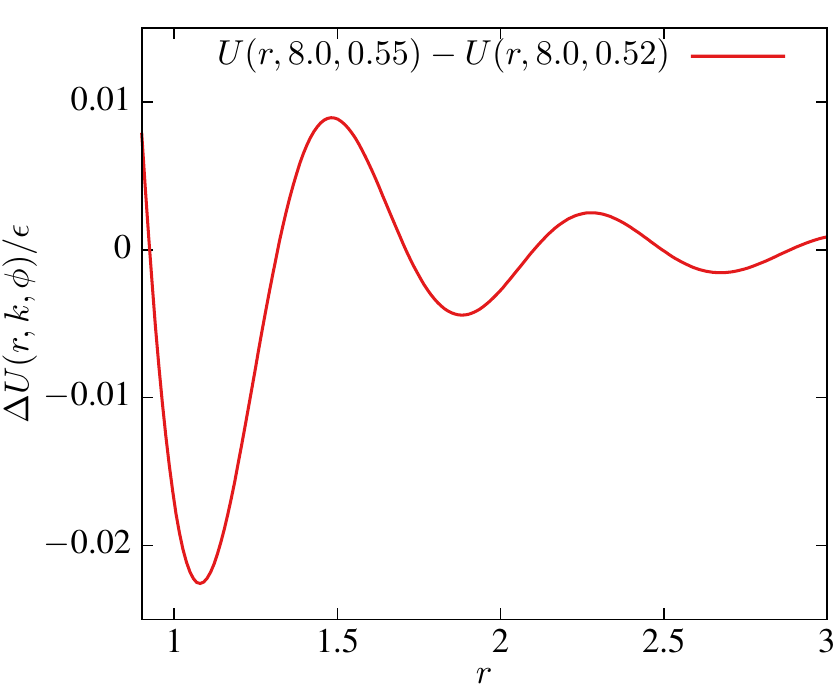}
  \end{center}
  \caption{
    The variation in the oscillating pair interaction potential with $\phi$ at
    fixed $k=8.0$ is small, yet leads to substantial gradients in the free
    energy (Fig.\ 4a, main text), and detectable differences in average
    potential energy per particle (Fig.\ 4b, main text), that are not
    discernible from the pattern registration depicted in Fig.\
    \ref{fullstack}a.
    \label{dufig}
  }
\end{figure}

\section{Toy Model: Lennard-Jones-Gauss Potentials in 2D}
As a consistency check, and as a non-trivial check on our analysis routines, we
apply them to a system for which we can determine optimal microscopic parameters
for the a given macroscopic state through direct calculations.

We study the effects of a one-parameter deformation of the relative
position of potential minima in 2D Lennard-Jones-Gauss systems
\cite{michaelljg} \via\ molecular dynamics simulations with HOOMD-Blue
\cite{hoomdblue}. The potential is given by
\begin{equation}\label{VLJG}
  V_\mathrm{LJG} = \varepsilon_\mathrm{LJ}
  \left(\left(\frac{\sigma_\mathrm{LJ}}{r}\right)^{12}
  -2\left(\frac{\sigma_\mathrm{LJ}}{r}\right)^{6}\right)
  -\varepsilon_\mathrm{G}e^{-(r-r_0)^2/2\sigma_\mathrm{G}^2}
\end{equation}
The phase diagram of this system has been previously determined in
\cite{michaelljg,carolynljg}. We use potential parameters as in prior published
work \cite{carolynljg}: 
$\varepsilon_\mathrm{G}=\tfrac34\varepsilon_\mathrm{LJ}$ and
$\sigma_\mathrm{G}=\tfrac{\sqrt{2}}{10}\sigma_\mathrm{LJ}$, and we work at
$T=\varepsilon_\mathrm{LJ}/10$. Finally, we define $\alpha\equiv
r_0/\sigma_\mathrm{LJ}$, and take it to be in the appropriate range to
self-assemble a square lattice.

We studied the constitutive relation between $\mu$ and $\alpha\equiv
r_0/\sigma_\mathrm{LJ}$ for the Lennard-Jones-Gauss system in 2D. In Fig.\
\ref{ljgalch} we plot the alchemical potential for a system of $N=1024$
Lennard-Jones-Gauss particles at $T=0.1\varepsilon_\mathrm{LJ}$ for $1.28125\le
\alpha\le1.46875$ where we observed the formation of a square lattice.
To determine the root of the constitutive equation, we performed a linear fit
\begin{equation}
  \frac{\mu}{\varepsilon_\mathrm{LJ}} \propto (\alpha-\alpha^*) \; ,
\end{equation}
where we find $\alpha^* = 1.383 \pm 0.001$.
For a square lattice, the ratio of the distance between
first neighbours and second neighbours is $\sqrt{2}$. We numerically computed
the locations of the minima of the Lennard-Jones-Gauss potential for $r_0$ in
the range where we observed formation of the square lattice, and found that
ratio of the second minimum $r_*^{(2)}$ to the first minimum $r_*^{(1)}$ was
$\sqrt{2}$ when $\alpha\approx1.4267$, which is well above the optimal value
indicated by the alchemical potential calculation. Because the temperature is
low, we expect the free energy to be dominated by the potential energy, so we
computed the average potential energy, and found that there was a potential
energy minimum between $1.359375\le \alpha\le1.390625$ in agreement with our
alchemical potential calculations.

Our alchemical potential calculation for the oscillating pair potential showed
that it is possible to detect effects that are not readily apparent by examining
the pattern registration between $g(r)$ and the pair potential, as shown in
Fig.\ \ref{fullstack}. However, a surprising result of the computations for the
Lennard-Jones-Gauss system in 2D is that even in simple cases where pattern
registration effects are discernible, the optimal pattern registration is not
what one would anticipate from na\"ive guessing.

In the Lennard-Jones-Gauss system, computations were performed at relatively low
temperatures, $T=0.1\varepsilon_\mathrm{LJ}$, and we would expect that at
sufficiently low temperatures, the free energy of the system is dominated by the
potential energy. In this case, we expect that the optimal $\alpha^*$ we've
determined above coincides with the value of $\alpha$ with the lowest ground
state energy.  For the range of $\alpha$ that self-assemble the square lattice,
we compute the ground state energy as a function of both $\alpha$ and the
lattice spacing. For each $\alpha$ we found the lattice spacing with the lowest
energy to get the ground state energy of the square lattice as a function of
$\alpha$. From this curve, we found the value of $\alpha$ with the minimum
ground state energy to be $\alpha=1.38342$, which accords very well with the
$\alpha^*$ we computed using the alchemical technique at
$T=0.1\varepsilon_\mathrm{LJ}$ (\cf\ Fig.\ \ref{ljgalch}, top panel and middle
panel). However, we note that Figs.\ \ref{r0less} and \ref{r0great} show the
surprising result that neither result corresponds with the na\"ive ansatz of
$\alpha_A=1.4267$ which that comes from fixing the relative distance between the
first and second minima of the Lennard-Jones-Gauss potential
$r_*^{(2)}/r_*^{(1)}=\sqrt{2}$, which we might expect to optimal because it
coincides with the appropriate distances for the square lattice in 2D (see Fig.\
\ref{ljgalch}, lower panel). It is possible that this discordance between the
na\"ive ansatz from pattern registration considerations, and the
thermodynamically optimal pair potential might differ in other systems, which
could be an important consideration in DNA-mediated nano-particle superlattice
assembly.\cite{park08,jonesetal,mirkindnasuper2012,mirkindnasuperrecon,dnaprogatom,
kaylieadvmat,mirkindnadesign,mirkindnaaniso,mirkinsci,gangdnaswitch,gangdnasuper}

\section{Oscillating Pair Potential}
\subsection{Simulation Protocol}
To evaluate the alchemical potentials conjugate to $k$ and $\phi$ for the
oscillating pair potential system \cite{opp}, we performed MD simulations of
$N=4069$ particles using HOOMD-Blue \cite{hoomdblue}, using a tabulated potential to
directly reproduce the simulation technique employed in \cite{opp}. The system
size of $N=4096$ particles was chosen so that sufficiently large changes in the
$k$ and $\phi$ parameters of the potential could be made, and still
replicate the simulation protocol followed in \cite{opp}. First, at each
state point $(k,\phi)$ we performed $NVT$ simulations with a cooling schedule
that was linear in temperature from an initial temperature of $3\epsilon$ to
$0.525\epsilon$ over $5\times10^7$ time steps, and then further from
$0.525\epsilon$ to $0.25\epsilon$ over $5\times10^7$ more time steps, to reach a
supercooled fluid. From the supercooled fluid snapshots, we launched several
$NVT$ simulations to nucleate the quasicrystal, in each case seeding the random
number generator of the Brownian integrator with a different integer. As a
consistency check we determined that on the boundaries of the stable range for
the intermediate density quasicrystal we observed a substantial fraction of
events in which we observed the nucleation of structures consistent with low
density or high density quasicrystal where appropriate, which suggests that due
to Lyapunov instabilities, our procedure leads to uncorrelated bulk structures
over the whole range. We allowed each simulation to run for up to $3\times 10^8$
time steps, checking the potential energy every $10^6$ time steps. Based on
empirical criterion of $U/N\epsilon < 0.25$ we determined that the quasicrystal
had nucleated or was about to nucleate, and then ran the simulation for a
further $5\times10^6$ time steps. For each putative nucleated quasicrystal, we
examined the structure and compared it with the structures reported in
\cite{opp}. Note that, likely due to the relatively small number of particles
($N=4096$) we found the region of self-assembly of the intermediate density
quasicrystal was smaller that that reported based on simulations of larger
systems in \cite{opp}. For each intermediate density quasicrystal, we
equilibrated for $1.3\times10^8$ time steps, and then over a period of $1.2\times
10^8$ time steps we stored a snapshot of the system every $4\times10^6$
time steps. All of the above time scales were determined to ensure decorrelation
based on autocorrelation measurements of the potential energy, and observing
diffusion of the quasicrystal in the simulation box. From these many independent
samples of the structure at a given $(k,\phi)$, we randomly selected snapshots
to re-equilibrate at nearby $(k',\phi)$ or $(k,\phi')$ according the finite
differencing scheme described above, which we did over $10^7$ time steps, again
chosen to ensure statistical independence based on measurement of potential
energy correlation. We then repeatedly sampled the potential energy every $10^5$
time steps over a further $10^7$ simulation time steps in order to estimate the
probability of making ghost Monte-Carlo moves between different values of $k$
and $\phi$ as described in detail above. At each nearby value of $(k,\phi)$ we
obtained several independent estimates of the transition probabilities, and
estimated the error from the standard deviation of the distribution of the
independent estimates.

\subsection{Pattern Registration in the Oscillating Pair Potential}
We computed the radial distribution function $g(r)$ for equilibrated snapshots
of the oscillating pair potential system at $k_\text{B}T=0.25\epsilon$ and
various values of $k$ and $\phi$ and compared it with the pair interaction
potential $U(r)$ in Fig.\ \ref{fullstack} to determine whether the decrease in
free energy we found in the alchemical potential calculation (Fig.\ 4a, main
text) could be detected in the pattern registration. Fig.\ \ref{fullstack} shows
no clearly discernible difference in the pattern registration. Furthermore, in
Fig.\ \ref{dufig} we plot potential energy difference as a function of $r$ at
fixed $k=8.0$ between $\phi=0.55$ and $\phi=0.52$, and see that over the range
of the oscillating pair potential, the differences are less than $3\%$ of
$\epsilon$.

\end{document}